\documentclass[prd, pdflatex, preprint, superscriptaddress, notitlepage, floatfix, nofootinbib]{revtex4-2}
\usepackage{xcolor}
\usepackage{graphicx}
\usepackage{wrapfig}
\usepackage[section]{placeins} 
\usepackage{amsmath}
\usepackage{amssymb}
\usepackage{multirow}
\usepackage{slashed}
\usepackage{physics}
\usepackage{bbm} 
\usepackage[colorlinks=true, linkcolor=blue, citecolor=blue, urlcolor=blue]{hyperref}
\usepackage{booktabs}
\usepackage{comment}
\usepackage{adjustbox}
\usepackage[normalem]{ulem}
\usepackage{etoolbox}
\usepackage{tocloft}
\usepackage{bbold}
\usepackage{array}
\usepackage{multirow}
\usepackage{amsmath}  
\usepackage{mathtools}

\usepackage{lineno}

\newcommand*\patchAmsMathEnvironmentForLineno[1]{%
	\expandafter\let\csname old#1\expandafter\endcsname\csname #1\endcsname
	\expandafter\let\csname oldend#1\expandafter\endcsname\csname end#1\endcsname
	\renewenvironment{#1}%
	{\linenomath\csname old#1\endcsname}%
	{\csname oldend#1\endcsname\endlinenomath}%
}

\AtBeginDocument{%
	\patchAmsMathEnvironmentForLineno{equation}%
	\patchAmsMathEnvironmentForLineno{align}%
	\patchAmsMathEnvironmentForLineno{flalign}%
	\patchAmsMathEnvironmentForLineno{alignat}%
	\patchAmsMathEnvironmentForLineno{gather}%
	\patchAmsMathEnvironmentForLineno{multline}%
}

\newcommand{\mres}{m_{\mathrm{res}}}

\newcommand{\barpsi}{\langle\bar{\psi}\psi\rangle}

\newcommand{\ua}{U(1)_{\mathrm A}}
\newcommand{\mq}{m_q}

\newcommand{\tc}{T_c}

\newcommand{\cdl}{\chi_\delta}
\newcommand{\cdsc}{\chi_\mathrm{disc}}

\newcommand{\pbp}{\langle \bar{\psi}\psi\rangle}

\newcommand{\cf}{\textit{cf.}}

\begin{document}
	

%
\preprint{KEK-CP-0412}

\title{The QCD phase diagram for three-flavor M\"obius domain-wall fermions}
\author{Yu Zhang}
\email{yzhang@physik.uni-bielefeld.de}
\affiliation{ Fakult\"{a}t f\"{u}r Physik, Universit\"{a}t Bielefeld, D-33615 Bielefeld, Germany}
 %
\author{Yasumichi Aoki}
\affiliation{RIKEN Center for Computational Science,  7-1-26
	\\	Minatojima-minami-machi, Chuo-ku, Kobe, Hyogo 650-0047, Japan}
 \author{Jishnu Goswami}
\affiliation{ Fakult\"{a}t f\"{u}r Physik, Universit\"{a}t Bielefeld, D-33615 Bielefeld, Germany}
\author{Shoji Hashimoto}
\affiliation{High Energy Accelerator Research Organization (KEK), Tsukuba 305-0801, Japan}
\affiliation{School of High Energy Accelerator Science, The Graduate University for Advanced Studies (Sokendai), Tsukuba 305-0801, Japan}
\author{Issaku Kanamori}
\affiliation{RIKEN Center for Computational Science,  7-1-26
	\\	Minatojima-minami-machi, Chuo-ku, Kobe, Hyogo 650-0047, Japan}
\author{Takashi Kaneko}
\affiliation{High Energy Accelerator Research Organization (KEK), Tsukuba 305-0801, Japan}
\affiliation{School of High Energy Accelerator Science, The Graduate University for Advanced Studies (Sokendai), Tsukuba 305-0801, Japan}
%
\author{Yoshifumi Nakamura}
\affiliation{RIKEN Center for Computational Science,  7-1-26
	\\	Minatojima-minami-machi, Chuo-ku, Kobe, Hyogo 650-0047, Japan}

\date{\today}

\begin{abstract}
  %
 We investigate the phase transition of Quantum Chromodynamics (QCD) with three degenerate quark flavors at zero baryon chemical potential. Using M\"{o}bius domain-wall fermions as the lattice fermion formulation, we ensure excellent chiral symmetry preservation. Our simulations are performed at three different temporal lattice extents, $N_{t}=6,8,12$, with a fixed lattice spacing $a=0.1361(20)$~fm, corresponding to temperatures of 242(4), 181(3), and 121(2) MeV, respectively. 
 We explore a range of quark masses and spatial volumes with aspect ratios $N_{s}/N_{t}$ spanning from 2 to 4. By analyzing the mass and volume dependencies of the plaquette, plaquette susceptibility, chiral condensate, chiral susceptibilities, and Binder cumulant, we identify the pseudocritical transition quark masses from our largest lattice volumes. For $N_t=6$, this is 184(10) MeV (determined from the plaquette susceptibility), corresponding to
 $m_\pi\simeq0.96$~GeV. For $N_t=8$ and 12, the transition points vary slightly depending on whether the total or disconnected chiral susceptibility is used, yielding ranges of $36(1)\text{--}39.1(9)$~MeV and $3.5(3)\text{--}3.7(2)$~MeV, respectively, in the $\overline{\text{MS}}$ scheme at a scale of $\mu=2$ GeV. In terms of the pseudoscalar mass, these correspond to
 $m_\pi\simeq427$--$445$~MeV and $133$--$137$~MeV. The negligible volume dependence at $N_t=6$ and 8, combined with finite-size scaling analysis at $N_t=12$ revealing volume growth significantly weaker than expected for a first- or second-order phase transition, points to a continuous crossover at these specific quark mass points. Additionally, we study the effects of residual chiral symmetry breaking on the chiral condensate and chiral susceptibilities 
   using two different values of $L_s$.

\end{abstract}
\date{\today}

\maketitle



\section{Introduction}
The nature of the finite-temperature QCD phase transition at zero baryon chemical potential depends on the number of quark flavors and their masses, a dependence summarized classically in the Columbia plot~\cite{Brown:1990ev}. At physical quark masses, it is well established that the transition is an analytic crossover~\cite{Aoki:2006we}, occurring at a pseudocritical temperature $T_{pc}$ of approximately $156$~MeV~\cite{Bazavov:2011nk, Bhattacharya:2014ara, HotQCD:2018pds, Borsanyi:2020fev}. The order of the transition in the various limiting corners of the Columbia plot, by contrast, has remained a long-standing and challenging problem for decades. This work addresses the corner of three degenerate light quarks ($N_f=3$), located in the bottom-left of the plot.

It is useful first to recall the two-flavor case ($N_f=2$), which introduces the concept central to what follows. Here, the order of the chiral transition is governed by the strength of $\ua$ symmetry breaking at the critical temperature $\tc$~\cite{Pisarski:1983ms}. If $\ua$ remains significantly broken, the transition is expected to be second order, in the $O(4)$ universality class; if $\ua$ is effectively restored, it could instead be first order or second order in the $U(2)_L \times U(2)_R \to U(2)_V$ universality class~\cite{Pisarski:1983ms, Butti:2003nu, Pelissetto:2013hqa, Grahl:2014fna}. Determining which scenario is realized in QCD has remained elusive due to the severe challenges of performing continuum and chiral extrapolations near the critical temperature in lattice QCD calculations~\cite{Ohno:2012br, HotQCD:2012vvd, Cossu:2013uua, Chiu:2013wwa, Bhattacharya:2014ara, Dick:2015twa, Tomiya:2016jwr, Brandt:2016daq, Ding:2020eql, Aoki:2020noz, Suzuki:2020rla, Aoki:2021qws, Kaczmarek:2021ser}. Recent lattice calculations have begun to address the fate of $\ua$ directly. A study using the Highly Improved Staggered Quark (HISQ) action in $N_f=2+1$ QCD found that $\ua$ remains effectively broken even after continuum and chiral extrapolations at $1.6\,\tc$, where $\tc\simeq 132$~MeV~\cite{Ding:2019prx, Kotov:2021rah}, pointing to a second-order $O(4)$ transition in the chiral limit at the physical strange quark mass~\cite{Ding:2020xlj}. A subsequent HISQ study confirmed that $\ua$ stays broken at $\tc$ in the chiral and continuum limits of $N_f=2+1$ QCD~\cite{Kaczmarek:2023bxb}. Another study using M\"{o}bius domain-wall fermions observed a strong suppression of the $\ua$ susceptibility in the chiral limit at temperatures of $153$--$204$~MeV, while it remained nonzero at $136$~MeV~\cite{JLQCD:2024xey}.

For the $N_f=3$ chiral limit\textemdash the subject of this work\textemdash the theoretical predictions themselves are in conflict. The universality argument of Pisarski and Wilczek~\cite{Pisarski:1983ms} predicts a first-order transition. Away from the chiral limit, explicit chiral symmetry breaking weakens this first-order transition, which eventually terminates at a second-order critical line in the three-dimensional $Z(2)$ universality class. The location of this critical line is of particular importance: its extension into the plane of nonzero baryon chemical potential determines the curvature of the $Z(2)$ critical surface and thereby influences the location\textemdash or even the existence\textemdash of the critical endpoint in the $T$--$\mu_B$ QCD phase diagram at the physical point, a central target of both theoretical~\cite{Fodor:2001pe, Fodor:2004nz, deForcrand:2006pv, Datta:2016ukp, Bazavov:2017dus, HotQCD:2017qwq, Borsanyi:2020fev} and experimental~\cite{STAR:2022hbp} efforts.

The first-order prediction has, however, been increasingly challenged. While the $\epsilon$-expansion of RG flows in $d=4-\epsilon$ dimensions~\cite{Pisarski:1983ms} and early $d=3$ analyses\textemdash both perturbative~\cite{Butti:2003nu} and via the nonperturbative functional renormalization group (FRG)~\cite{Resch:2017vjs}\textemdash supported a first-order transition (no infrared-stable fixed point being found, regardless of the status of $\ua$), a renewed $d=3$ FRG treatment that includes all relevant operators~\cite{Fejos:2022mso, Fejos:2024bgl}, together with conformal bootstrap studies~\cite{Kousvos:2022ewl}, has identified an infrared-stable fixed point. This suggests a possible second-order transition, provided $\ua$ is effectively restored at the critical point. Recent conjectures~\cite{Pisarski:2024esv} and extended linear sigma model analyses~\cite{Giacosa:2024orp} further emphasize that a vanishing 't Hooft determinant is necessary for a second-order transition in massless $N_f=3$ QCD. These findings collectively point to the possibility of a second-order transition in the $m_{u,d}=0$ limit for all values of the strange quark mass, though the universality class differs between the $N_f=2$ and $N_f=3$ chiral limits; in these frameworks, the fate of $\ua$ plays a crucial role in determining the order of the phase transition and the universality class for $N_f=3$ massless QCD. Adding to the complexity, a truncated Dyson--Schwinger study predicts a second-order transition in the $N_f=3$ chiral limit irrespective of the fate of $\ua$ at $\tc$~\cite{Bernhardt:2023hpr}. In addition, recent theoretical work revisited the QCD chiral phase transition by emphasizing additional 't Hooft anomaly constraints. It argued that, if the chiral phase transition is second order or very weakly first order for $N_f \geq 2$, the conventional Ginzburg-Landau picture is strongly constrained, and a beyond-Ginzburg-Landau conformal-manifold scenario may be a viable possibility, particularly for $N_f \geq 3$~\cite{Chen:2026jla}. While these theoretical approaches provide valuable qualitative insight, their predictions do not agree.

The key open questions can therefore be stated plainly. Does a first-order region exist in the $N_f=3$ chiral limit? If not, what is the corresponding universality class? If it does exist, what is the critical mass separating the first-order and crossover regions? Answering these questions requires nonperturbative lattice QCD simulations.

Pinpointing the transition in the $N_f=3$ chiral region on the lattice is difficult, however, owing to the high computational cost at small quark masses and the sensitivity to cutoff effects. Early studies with staggered~\cite{JLQCD:1998mja, Liao:2001en, Karsch:2001nf, deForcrand:2003vyj} and Wilson fermions~\cite{Iwasaki:1995ij}, as well as $\mathcal{O}(a)$-improved Wilson fermions~\cite{Jin:2014hea}, identified a first-order chiral phase transition region on coarse lattices. Subsequent studies, however, revealed that the critical pion mass extracted from such coarse lattices depends heavily on both the lattice action and the lattice spacing~\cite{Karsch:2003va, deForcrand:2006pv, deForcrand:2007rq, Endrodi:2007gc, Ding:2011du, Varnhorst:2015lea, Bazavov:2017xul, Jin:2017jjp, Kuramashi:2020meg}: for both formulations the critical pion mass decreases as the lattice spacing is reduced. Moreover, a systematic discrepancy persists between them, with Wilson-type fermions consistently yielding larger critical pion masses than staggered fermions\textemdash a difference likely due to the substantial chiral symmetry breaking inherent in Wilson fermions. A similar trend appears in $N_f=4$ QCD, where the critical pion mass shows strong cutoff effects for both staggered~\cite{deForcrand:2017cgb} and Wilson fermions~\cite{Ohno:2018gcx}, even approaching zero as the lattice spacing decreases. More recently, an $\mathcal{O}(a)$-improved Wilson study placed an upper bound $m_\pi^c \lesssim 110$~MeV in the continuum limit of $N_f=3$ QCD~\cite{Kuramashi:2020meg}, a tricritical-scaling analysis with unimproved staggered fermions found evidence of a second-order transition in the chiral and continuum limits\textemdash the same conclusion was reached when 
the analysis of Ref.~\cite{Cuteri:2021ikv}
was applied to the improved Wilson data of~\cite{Kuramashi:2020meg}
\textemdash and a HISQ study likewise suggested a second-order transition in the chiral limit~\cite{Dini:2021hug}. These results leave little room for a first-order region, though they do not rule it out entirely. Crucially, all existing results come from staggered or Wilson-type fermions, which break chiral symmetry partially or completely, with full restoration only in the continuum limit. This underscores the need to explore the $N_f=3$ chiral phase transition using a chiral fermion formulation in order to achieve a more comprehensive understanding.

In this work, we present the first investigation of the finite-temperature phase transition in $N_f=3$ QCD using M\"{o}bius domain wall fermions~\cite{Brower:2012vk}, a generalization of the Shamir domain wall fermion formulation. This formulation precisely preserves chiral symmetry at finite lattice spacing when the fifth dimension is sufficiently large, while significantly reducing residual chiral symmetry breaking even for smaller extents than Shamir domain wall fermions. Preliminary results from this work have been reported in previous lattice conference proceedings~\cite{Nakamura:2022abk, Zhang:2022kzb, Zhang:2024ldl, Zhang:2025vns}. The paper is organized as follows. In Sec.~\ref{label3}, we introduce the observables used for determining the finite-temperature phase transition in $N_f=3$ QCD. In Sec.~\ref{label4}, we provide details of the simulation parameters. Sections~\ref{sec:zeroT} and \ref{sec:finiteT} present our numerical results
at zero and finite temperature, respectively. Finally, we summarize our findings and discuss future prospects in Sec.~\ref{label5}. Additional details on the M\"{o}bius domain-wall fermion action and the axial Ward--Takahashi identity are collected in Appendix~\ref{sec:appendix_action}, simulation parameters are compiled in Appendix~\ref{app:params}, the rational function reweighting procedure is described in Appendix~\ref{app:reweight}, and additional residual mass and pion screening mass results are presented in Appendix~\ref{app:additional mass}.


\emph{}
%


\section{Observables}
\label{label3}
The M\"obius domain-wall fermion action used in this work is detailed in Appendix~\ref{sec:appendix_action}. Based on this action, we compute several key physical quantities in this section to quantify the residual chiral symmetry breaking and the nature of the QCD phase transition.
\subsection{Residual mass}

The M\"obius domain-wall fermion formulation with a finite fifth dimension $L_s$ allows mixing between the left- and right-handed modes, leading to residual chiral symmetry breaking. At leading order 
 in the lattice spacing expansion, this breaking is characterized by a mass-independent residual mass $\mres$, which acts as an additive renormalization to the bare quark mass, giving the total quark mass $m = \mq + \mres$. In practice, $\mres$ is determined non-perturbatively from the Ward-Takahashi identity of the axial current, as shown in Eq.~\eqref{eq:WTI} in Appendix~\ref{sec:appendix_action}.
In this identity, $m_q$ denotes the input bare quark mass, and $J_5^a$  represents the pseudoscalar density constructed from the quark fields on the boundaries of the fifth dimension, given by
\begin{equation}
	\label{eq:J_5}
	J_5^{a}(x) = \bar \Psi_{x, 0} \lambda^a D_{-}^{(0)}P_+ \Psi_{x, L_s-1} - \bar\Psi_{x,L_s-1} \lambda^a D_{-}^{(L_s-1)}P_{-} \Psi_{x,0}   \,,
\end{equation}
where $P_\pm = \frac{1}{2}(1\pm \gamma_5)$ are the chiral projection operators. 

An additional current, the mid-point current $J_{5q}^a$, is built from the quark fields at $L_s/2$ and $L_s/2-1$:
 \begin{equation}
 	\label{eq:J_5q}
 	J^{a}_{5q}(x)= \bar\Psi_{x, L_s/2}\lambda^a D_{-}^{(L_s/2)} P_+ \Psi_{x, L_s/2-1}  - \bar \Psi_{x, L_s/2-1}  \lambda^a D_{-}^{(L_s/2-1)}P_{-} \Psi_{x, L_s/2} \,.
 \end{equation}
The mid-point current $J^a_{5q}$ represents an additional contribution to the continuum Ward-Takahashi identity,
which is recovered only in the $L_s \to \infty$  limit for the non-singlet flavor. The residual mass $\mres$ can be determined from this mid-point current.

At large source-sink separations $t$, the ratio of the mid-point correlation function to the pion correlation function, 
\begin{equation}
	\label{eq:ratio}
	R(t) = \frac{\left\langle \sum_{\vec x}J_{5q}^{a}(\vec{x},t)\,J_5^{a}(\vec{0},0) \right\rangle}{\left\langle \sum_{\vec x}J_5^a(\vec{x},t)\,J_5^a(\vec{0},0)\right\rangle} \,,
\end{equation}
converges to a plateau value, denoted as $\mres(m_q)$, which depends weakly on $m_q$. This mass dependence is a lattice artifact arising from higher-dimension operators in the axial current divergence and represents an $\mathcal{O}(a^2)$ contribution~\cite{RBC:2006jmm,RBC:2007yjf}. 

The above determination ensures that $m_\pi$ vanishes when $m_q + m_{\rm res}(m_q) = 0$. The chiral limit is defined by vanishing total quark mass, $m_q + m_{\rm res} = 0$, or equivalently $m_q = -m_{\rm res}$, where $m_{\rm res}$ denotes a mass-independent constant. To obtain this constant, a natural approach is to extrapolate $m_{\rm res}(m_q)$ to the chiral limit, yielding a value denoted as $m_{\rm res}(-m_{\rm res})$. Another popular choice is to use $m_{\rm res}(0)$. We will use the latter throughout this work.

\subsection{Chiral condensate, chiral susceptibilities, and Binder cumulant }
The chiral condensate $\langle \bar{\psi}{\psi} \rangle$ serves as the order parameter for the QCD chiral phase transition, with its value being zero in the chirally symmetric phase and nonzero in the chirally broken phase.  In the M\"obius domain wall formulation, the physical 4D quark fields, $\psi$ and $\bar{\psi}$, are constructed from the chiral modes localized on opposite boundaries of the fifth dimension as follows: 
\begin{equation}
	\label{eq:q_x}
	%
	\psi_x  = P_{-} \Psi_{x,0} + P_{+} \Psi_{x, L_s-1} = [\mathcal{P}^{\dagger}\Psi]_{x,0}\,,
\end{equation}
\begin{equation}
	\label{eq:qbarx}
	%
	\bar{\psi}_x =  \bar\Psi_{x, 0} (-D_{-}^{(0)}) P_{+} +   \bar{\Psi}_{x,L_s-1}(- D_{-}^{(L_s - 1)})P_{-} = [\bar{\Psi}(-D_{-})\mathcal{P}^{\dagger}]_{x,L_s-1}  \,.
\end{equation}
Here, $\mathcal{P}$ is a permutation matrix defined as $\mathcal{P}_{ss^{\prime}}  = \delta_{ss^{\prime}} P_{-} + \delta_{s^{\prime}, (s+1) \,\bmod\, L_s} P_{+}$. The quark mass $m_q$ enters the fermion action through the coupling
between the boundaries of the fifth dimension, as shown in the last two terms of Eq.~\eqref{eq:MDWF}. Here, $D^{(s)}_{-}$ is related to the Wilson Dirac matrix operator (for details, please see Appendix~\ref{sec:appendix_action}).

To calculate the fermionic observables, we first consider the quark propagator in a fixed gauge background $U$. In the fermionic path integral formalism, it is given by 
\begin{align}\label{eq:propagator}
	\langle \psi_x\bar{\psi}_y\rangle_{F} &= \frac{1}{Z_F(U)} \int \mathcal{D}\Psi \mathcal{D}\bar{\Psi}\mathcal{D}\Phi \mathcal{D}\bar{\Phi}\, \psi_x\bar{\psi}_y \,e^{ - \bar{\Psi}D_{\text{MDWF}}(m_q, U)\Psi-\bar{\Phi}D_{\text{MDWF}}(1,U)\Phi}\,, 
\end{align} 
where $Z_F(U)$ is the fermionic partition function in the fixed gauge background $U$, and $\Phi_{x, s}$,$\bar{\Phi}_{x, s}$ are the 5D bosonic Pauli-Villars fields. Integrating out the fermion and Pauli-Villars fields yields
\begin{align}
		\langle \psi_x\bar{\psi}_y\rangle_{F}  = \frac{1}{1-m_q} \left( \left[D^{L_s}_{\text{ov}}(m_q) \right]^{-1}_{x,y} - \delta_{x,y} \right)\,. 
\end{align}
For notational convenience, we define the operator $\tilde{D}^{L_s}_{\text{ov}}(m_q)$ by
\begin{align}
	 \left[ \tilde{D}^{L_s}_{\text{ov}}(m_q) \right]^{-1}_{x,y} = \frac{1}{1-m_q} \left( \left[D^{L_s}_{\text{ov}}(m_q) \right]^{-1}_{x,y} - \delta_{x,y} \right)\,. 
\end{align}	
$D^{L_s}_{\mathrm{ov}}$ is the 
4D effective overlap operator utilizing the polar 
approximation to the sign function, given by~\cite{Brower:2012vk}:
\begin{align}\label{eq:dov}
		D^{L_s}_{\mathrm{ov}}(m_q) = \frac{1+m_q}{2} + 	\frac{1-m_q}{2}\,\gamma_5\,\epsilon_{L_s} [H] \quad \text{with}\quad \epsilon_{L_s} [H] = \frac{(1+H)^{L_s} - (1-H)^{L_s}}{(1+H)^{L_s} + (1-H)^{L_s}}
\end{align}	
	Here, $H= \gamma_5 D^{\text{M\"obius}}(M_5)$, with $D^{\text{M\"obius}}(M_5)$ being the M\"obius kernel, whose explicit form is given in Eq.~\eqref{eq:mobius} of Appendix~\ref{sec:appendix_action}. In the limit $L_s \to \infty$, $\epsilon_{L_s} [H]$ converges to $\text{sign}(H)$, recovering the exact overlap operator. The exact mapping between the 4D effective operators and the 5D M\"obius domain wall fermion operators is given by:
\begin{align}\label{eq:connect_5D_MDWF}
		D^{L_s}_{\mathrm{ov}}(m_q) =  \mathcal{P}^{\dagger} D^{-1}_{\text{MDWF}}(1)  D_{\text{MDWF}}(m_q) \mathcal{P} 
\end{align}
	The corresponding 4D quark propagator is then obtained from the inverse of the 5D Dirac operator through the relation:
\begin{align}\label{eq:physical_prop_MDWF}
 \left[ \tilde{D}^{L_s}_{\text{ov}}(m_q) \right]^{-1} = \mathcal{P}^{\dagger} D^{-1}_{\text{MDWF}}(m_q) (-D_{-}) \mathcal{P}^{\dagger}
\end{align}	
where $D^{-1}_{\text{MDWF}}(m_q) \equiv \langle \Psi  \bar{\Psi} \rangle$. It should be noted that, because directly inverting the 4D effective overlap operator is computationally expensive, we instead construct a 5D source vector, $\xi_{\text{MDWF},0} = -D_{-}\mathcal{P}^{\dagger}\xi_0$, from the 4D source $\xi_0$. We then invert the 5D M\"obius domain wall fermion Dirac operator by solving $G  = D^{-1}_{\text{MDWF}}(m_q) \xi_{\text{MDWF},0}$ using the even-odd preconditioned conjugate gradient method. Finally, the 5D solution $G$ is projected back to 4D via $\mathcal{P}^{\dagger}G$ to yield the physical propagator $[\tilde{D}^{L_s}_{\text{ov}}(m_q)]^{-1}_{x,y}$. 
 
The $N_f$-flavor M\"{o}bius domain wall fermion chiral condensate is then defined as
\begin{align}\label{eq:pbp}
\barpsi =  \frac{T}{V}\frac{\partial \ln Z}{\partial m_q} = \frac{N_f}{N_{s}^3N_{t}} 
\left\langle\mathrm{Tr}\,
\left(\tilde{D}^{L_s}_{\mathrm{ov}}(m_q)\right)^{-1} 
\right\rangle
\end{align}
To ensure that $\langle \bar{\psi}{\psi} \rangle$, as computed in Eq.~\eqref{eq:pbp}, is finite and well-defined in the continuum limit, both additive and multiplicative renormalizations have to be applied when the quark mass is finite. 
The dominant ultraviolet divergence, which necessitates an additive renormalization, is a power-law term of the form $C^D (m_q + x \mres) /a^2$, where $C^D$
	and $x = O(1)$ are a priori unknown coefficients~\cite{Sharpe:2007yd}. Here the $x \mres a^{-2}$ term arises from the finite extent $L_s$ of the fifth dimension:
the residual violation of chiral symmetry induces a similar effect to that of the quark mass term. A subleading logarithmic divergence 
also exists, but its contribution is highly suppressed and can be safely ignored~\cite{Bazavov:2011nk}. Therefore, the chiral condensate behaves as 
\begin{align}\label{eq:DWF_pbp}
\barpsi|_{\mathrm{DWF}} \sim C^D \frac{\mq + x\mres}{a^2} + \barpsi|_{\mathrm{cont}} + ...\,.
\end{align}
Since $x\ne1$, if we perform an extrapolation to the chiral limit $\mq+\mres=0$ 
as one usually does for low-energy physics, 
 a UV-divergent piece still remains and 
  behaves as $C^D \frac{(x-1)\mres}{a^2}$. This unwanted contribution can only be effectively controlled by increasing 
  		$L_s$. On the one hand, if we know the value of $C^D$ and $x$, we can remove the additive divergence part explicitly; on the other hand, 
  		because the UV divergence term $C^D \frac{\mq + x\mres}{a^2}$ is temperature independent, it can be canceled
  		by subtracting the zero-temperature condensate $\barpsi^{\mathrm{T}=0}$ from the finite-temperature one $\barpsi^{\mathrm{T}>0}$ at 
  		the same bare parameters. 
  		
To eliminate the remaining multiplicative divergence, 
we multiply by the scalar density renormalization factor
$Z_s^{\overline{\mathrm {MS}}}(\mu=2\, \mathrm{GeV})= 1/Z_m^{\overline{\mathrm {MS}}}(\mu=2\, \mathrm{GeV})$,  
where $Z_m^{\overline{\mathrm {MS}}}(\mu=2\, \mathrm{GeV})$ is the quark mass renormalization constant obtained 
from NNNLO running~\cite{Chetyrkin:1999pq}.
The two renormalized chiral condensates are then defined as:
\begin{align}\label{eq:renorm_pbp1}
  \frac{\barpsi^{\mathrm{T}>0} - \barpsi^{\mathrm{T}=0}}{Z_m^{\overline{\mathrm {MS}}}(\mu=2\,\mathrm{GeV})} = [\barpsi^{\mathrm{T}>0} - \barpsi^{\mathrm{T}=0} ]^{\overline{\mathrm {MS}}}(\mu=2\,\mathrm{GeV}) \,,
\end{align}
\begin{align}\label{eq:renorm_pbp2}
	 \frac{	\langle \bar{\psi} \psi \rangle } {Z_m^{\overline{\mathrm {MS}}}(\mu=2\,\mathrm{GeV})} 
	 - C^D \frac{m_q + x m_{res}}{a^2} 
	 Z_m^{\overline{\mathrm {MS}}}( \mu = 2\,\mathrm{GeV}) = [\langle \bar{\psi} \psi \rangle -  C^D \frac{m_q + x m_{res}}{a^2} ]^{\overline{\mathrm {MS}}}(\mu=2\,\mathrm{GeV})\,.
\end{align}
The total chiral susceptibility $\chi_{\rm{tot}}$ measures the response of the chiral condensate to a small change in the quark mass, and is defined as
\begin{align}\label{eq:chi_sus}
\chi_{{\rm tot}} = \frac{\partial \langle \bar \psi \psi \rangle }{\partial m_q}\,.
\end{align}
 It contains two parts: 
the quark-line disconnected part 
 $\chi_{\mathrm{disc}}$, which describes the fluctuations of the chiral condensate:
\begin{align}\label{eq:chi_disc}
\chi_{{\rm disc}} = \frac{N_f^2}{N_s^3 N_t} \left\{ \left\langle \left[ {\rm Tr} \left(\tilde{D}^{L_s}_{\mathrm{ov}}(m_q)\right)^{-1} \right]^2 \right\rangle - \left\langle\mathrm{Tr}\,
\left(\tilde{D}^{L_s}_{\mathrm{ov}}(m_q)\right)^{-1} \right\rangle^2 \right\}\,,
\end{align}
and the quark-line connected part $\chi_{\mathrm{con}}$ (also denoted $\chi_{\delta}$):
\begin{align}\label{eq:chi_con}
 \chi_{\delta} = -\frac{N_f}{N_s^3 N_t}   \left\langle {\rm Tr}\,\left[\left(\tilde{D}^{L_s}_{\mathrm{ov}}(m_q)\right)^{-2}\right]   \right\rangle.
\end{align}
Alternatively, $\chi_{\delta}$ can also be written in terms of the integrated two-point flavor non-singlet scalar correlator:
\begin{align}\label{eq:chi_con_corr}
	 \chi_{\delta} = - N_f\, \sum_{x} \left\langle \text{tr} \left[(\tilde{D}^{L_s}_{\mathrm{ov}})^{-1}(x,0)  ( \tilde{D}^{L_s}_{\mathrm{ov}} )^{-1}(0,x) \right]\right\rangle.
\end{align}	
Here, tr denotes a trace over spinor and color indices only, while Tr indicates a trace that additionally includes the space-time volume. We employ stochastic trace estimation to calculate the chiral condensate and its fluctuations. To determine the connected chiral susceptibility, we utilize the integrated two-point flavor non-singlet scalar correlator, which is readily available from the two-point correlator 
measurements performed for all gamma matrix insertions 
in the residual mass calculation.

The disconnected chiral susceptibility $\chi_{\mathrm{disc}}$ serves as  a sensitive probe of the chiral phase transition. In the chiral and infinite-volume limits, $\chi_{\mathrm{disc}}$ diverges at the critical point, whereas
$\chi_{{\rm \delta}}$ remains finite. At finite quark mass, this divergence is smoothed into a peak that reaches its maximum at the pseudocritical temperature. Unlike the connected part, $\chi_{\mathrm{disc}}$ does not exhibit an additive divergence but requires multiplicative renormalization~\cite{Bazavov:2011nk}. 
This is achieved using
 $Z_m^{\overline{\mathrm {MS}}}(\mu=2\,\mathrm{GeV})$ as follows:

\begin{align}\label{eq:renorm_chi_disc}
\chi_{\mathrm{disc}}^{\overline{\mathrm {MS}}}(\mu=2\,\mathrm{GeV}) = \frac{\chi_{{\rm disc}}}{\left(Z_m^{\overline{\mathrm {MS}}}(\mu=2\,\mathrm{GeV})\right)^2}\,.
\end{align}
The connected chiral susceptibility, originating from the explicit quark mass dependence of the chiral condensate, requires both additive and multiplicative renormalization. Its additive divergence, which behaves as $C^D/a^2$, can be removed either by subtracting the vacuum expectation value or, if $C^D$ is known, by explicit subtraction. The multiplicative renormalization can be 
addressed in the same manner as for $\chi_{\rm{disc}}$.

To determine the order of the phase transition, we measure the Binder
cumulant of the chiral condensate~\cite{cite-key}. For each gauge
configuration $U$ we construct the fermion-integrated condensate
\begin{align}\label{eq:MU}
	M(U) = \frac{N_f}{N_s^3N_t}\,
	\mathrm{Tr}\left[\tilde{D}^{L_s}_{\mathrm{ov}}(\mq)\right]^{-1}(U),
\end{align}
the configuration-wise, c-number representative of $\bar\psi\psi$, so that
$\pbp=\langle M\rangle_U$ with $\langle\cdots\rangle_U$ the gauge-ensemble
average [\cf~Eq.~\eqref{eq:pbp}]. The Binder cumulant is then
\begin{align}\label{eq:B4}
	B_4(\bar\psi\psi)
	=\frac{\left\langle(\delta M)^4\right\rangle_U}
	{\left\langle(\delta M)^2\right\rangle_U^2},
	\qquad
	\delta M = M(U)-\langle M\rangle_U ,
\end{align}
i.e., the Binder cumulant of the gauge-ensemble probability distribution
$P(M)$ of the chiral order parameter, built from the quark-line
disconnected moments.\footnote{The cumulants of the microscopic condensate operator, given by
	$\partial^n\ln Z/\partial\mq^n$, contain in addition quark-line connected contributions; at second order this is the decomposition
	$\chi_{\rm tot}=\cdsc+\cdl$ of Eqs.~\eqref{eq:chi_disc} and~\eqref{eq:chi_con}. The connected contributions are subleading near the chiral critical point and regular away
	from it, so that the limiting values of $B_4(\bar\psi\psi)$ given below are
	the same for either definition.}

To compute $B_4(\bar\psi \psi)$ accurately, we use an unbiased stochastic estimation method with a set of $N=10$ independent $Z_2$ noise vectors. Letting $O_i = \frac{N_f}{N_{s}^3N_{t}}\xi_i^\dagger [\tilde{D}^{L_s}_{\text{ov}}(m_q)]^{-1} \xi_i$ be the estimate from a single noise vector, the required moments are constructed by averaging over all unique combinations of these individual estimators~\cite{deForcrand:2017cja}.
This approach is crucial for eliminating statistical bias that would arise from reusing noise vectors within a product. For a fixed gauge configuration, the unbiased stochastic estimators of the
first four powers of $M(U)$ are
\begin{align}
	\overline{M}   &= \frac{1}{\binom{10}{1}} \sum_{i=1}^{10} O_i ,\\
	\overline{M^2} &= \frac{1}{\binom{10}{2}} \sum_{1 \le i < j \le 10} O_i O_j ,\\
	\overline{M^3} &= \frac{1}{\binom{10}{3}} \sum_{1 \le i < j < k \le 10} O_i O_j O_k ,\\
	\overline{M^4} &= \frac{1}{\binom{10}{4}} \sum_{1 \le i < j < k < l \le 10} O_i O_j O_k O_l ,
\end{align}
where the overline denotes the average over the noise vectors on a fixed
gauge configuration and the indices $i,j,k,l$ label these vectors, not quark flavors. The
gauge-ensemble average is performed subsequently. We can distinguish different types of phase transitions by the result of $B_4(\bar\psi \psi)$. In the thermodynamic limit, $B_4(\bar\psi \psi)=1$ represents a first-order phase transition; $B_4(\bar\psi \psi)=3$ corresponds to a smooth crossover; and $B_4(\bar\psi \psi)= 1.604$ indicates a second-order phase transition belonging to the 3D $Z(2)$ universality class~\cite{Blote_1995}. However, at finite volume, $B_4(\bar\psi\psi)$ deviates from its 
infinite-volume limits ($B_4(\bar\psi\psi)=1$ for a first-order 
transition and $B_4(\bar\psi\psi)=3$ for a crossover) due to 
finite-volume corrections, approaching these values only in the 
thermodynamic limit. In contrast, at a second-order critical point, 
the $B_4(\bar\psi\psi)$ curves for different volumes are expected 
to intersect in the vicinity of the critical point for sufficiently 
large volumes. Our goal is to determine the order of the phase transition at the transition mass point for a fixed temperature via measurements of the Binder cumulant.
\section{Lattice Parameters and Simulation Details}
\label{label4}
We perform $N_f=3$ QCD simulations using the tree-level improved Symanzik gauge action and the M\"{o}bius domain-wall fermion action with three steps of stout smearing~\cite{Zhang:2022kzb,Nakamura:2022abk}. The simulations are carried out with the Grid code, using a version optimized for the A64FX architecture of the Fugaku supercomputer~\cite{Meyer:2019gbz}. We choose a fixed gauge coupling of $\beta=4.0$, 
corresponding to a lattice spacing of $a=0.1361(20)$ fm. The lattice spacing is determined from the Wilson-flow scale $t_0$, defined by the condition ${t^2\langle E(t) \rangle }|_{t=t_0} = 0.3$, with $\sqrt{t_0}$ set to its physical value of 0.1465(21)(13) fm as obtained in  $N_f=2+1$ QCD~\cite{Borsanyi:2012zs}. 
The calculations are carried out on lattices of size $N_{s}^3 \times N_t \times L_s = N_{s}^3 \times 6 \times 16$ with $N_{s} = 12, 16$, $N_{s}^3 \times 8 \times 16$ with $N_{s} = 16, 24$; as well as $N_{s}^3 \times 12 \times 16$ with $N_{s} = 24, 36$ and $48$. These temporal extents $N_t=6,8,12$
correspond to temperatures of 242(4), 181(3) and 121(2) MeV, respectively, and encompass a range of quark masses. Specifically, we simulated 41 and 21 quark masses within $am_q\in[0,0.4]$ and $[0.05,0.25]$ for the
$12^3\times6\times16$ and $16^3\times6\times16$ ensembles, respectively; 11 and 15 quark masses within $am_q\in[0,0.2]$ and $[0,0.14]$ for the
$16^3\times8\times16$ and $24^3\times8\times16$ ensembles, respectively; 26 quark masses within $am_q \in [-0.006, 0.1]$ for $24^3 \times 12\times 16$ ensembles, 7 quark masses within $am_q \in [-0.005, 0.001]$ for $36^3 \times 12 \times 16$ ensembles, and $am_q = -0.003$ and $-0.004$ for $48^3 \times 12\times 16$ ensembles. $L_s$ is 16 for all the above ensembles. To check the residual chiral symmetry breaking effect and the stability of the molecular dynamics evolution in the Hybrid Monte Carlo simulation for the above $N_t=12$, $L_s=16$ ensembles, we generated $24^3 \times 12 \times 32$ ensembles with 5 quark masses in the range between $-0.001$ and $0.003$. For each parameter set, we accumulate between $\sim10^4$ and $\sim10^5$
trajectories, performing measurements every 10 trajectories, except for the
$N_t=8$ ensembles, where configurations were analysed every 100 trajectories.
Further details are given in
Tables~\ref{sup:table_nt6}--\ref{sup:table_nt12} of Appendix~\ref{app:params}.

Apart from the finite temperature ensembles, we also generated zero-temperature ensembles at several gauge couplings: $\beta=4.0$ and $4.1$ on $24^3 \times 48 \times 16$ ensembles, and $\beta=4.17$ on $32^3 \times 64 \times 16$ ensembles, each 
with several quark masses (see Table~\ref{sup:table_zeroT} in Appendix~\ref{app:params}). These zero-temperature ensembles are utilized to calculate the lattice spacing, pion mass, residual mass, 
and the zero-temperature chiral condensate, which is used to subtract the UV divergence present in the finite-temperature chiral condensate. For the stochastic estimation of the chiral condensate, $N_{src}=10$ noise vectors were used per configuration. While the random seeds remained independent for every measurement within an individual stream, for ensembles with multiple streams, the seeds were identical across different streams for the same trajectory number. 
To ensure this correlation across streams did not introduce significant bias, we performed a cross-check on a representative ensemble using independent random seeds for all measurements. The results were found to be consistent within statistical errors. Since the configuration generation process utilized independent seeds and was not affected, we retain the original results to avoid additional computational cost.

Our M\"{o}bius domain-wall fermion configurations are generated using the same algorithmic setup as in $N_f=2+1$ QCD simulations implemented in the Grid code. In this setup, the two degenerate light flavors are simulated using the standard Hybrid Monte Carlo (HMC) algorithm, while the remaining single flavor is simulated using the Rational Hybrid Monte Carlo (RHMC) algorithm. In the present $N_f=3$ degenerate case, we set $m_s=m_l=m_q$, so the product of the two-flavor HMC determinant and the one-flavor RHMC determinant reproduces the desired $\det\left[D^{\dagger}_{\text{MDWF}}(m_q) D_{\text{MDWF}} (m_q)\right]^{3/2}$. After integrating out the Grassmann-valued fermions and Pauli-Villars fields, the fractional power appears only in the one-flavor RHMC factor. Introducing bosonic pseudofermion fields $\phi_l$ and $\phi_s$ for the two-flavor HMC and one-flavor RHMC sectors, respectively, the determinant ratio can be represented as:
\begin{align}\label{1flavor det_ratio}
\frac{\det \left[D^{\dagger}_{\text{MDWF}}(m_q) D_{\text{MDWF}} (m_q)\right]^{3/2}}{\det\left[D^{\dagger}_{\text{MDWF}}(1) D_{\text{MDWF}}(1)\right]^{3/2}} =\int \mathcal{D} \phi_l^{\dagger}  \mathcal{D} \phi_l \mathcal{D} \phi_s^{\dagger}  \mathcal{D} \phi_s  \,e^{ -S_{pf}}\,,
\end{align} 
where the pseudofermion action is given by 
\begin{align}
S_{pf} = \phi_l^{\dagger} \left(\frac{D^{\dagger}_{\text{MDWF}}(m_q) D_{\text{MDWF}} (m_q)}{D^{\dagger}_{\text{MDWF}}(1) D_{\text{MDWF}} (1)}\right)^{-1}   \phi_l + \phi_s^{\dagger} \left(\frac{D^{\dagger}_{\text{MDWF}}(m_q) D_{\text{MDWF}} (m_q)}{D^{\dagger}_{\text{MDWF}}(1) D_{\text{MDWF}} (1)}\right)^{-1/2}   \phi_s\,.
\end{align} 
 The fractional power in the one-flavor RHMC part is approximated using a rational function,  written in the partial fraction form $x^{-1/2} \simeq  {a_0 + \sum_{i=1}^{n} \frac{a_i}{x+b_i}}$. The multi-shift solver solves the set of linear systems $\left(D^{\dagger}_{\text{MDWF}}(m_q) D_{\text{MDWF}} (m_q) + b_i\right)v_i = \phi_s$ 
 for all shifts $b_i$ simultaneously, requiring essentially the same number of applications as for a single matrix inversion using the conjugate gradient method~\cite{Frommer:1995ik}. 

 In our $N_f=3$ simulations on $36^3\times 12 \times 16$ ensembles, the number of terms used in the rational function was found to be  insufficient. This limited precision causes the generated gauge configurations to deviate slightly from the target
  distribution. To correct for this, we employ a rational function reweighting procedure. Further details regarding this reweighting can be found in Appendix~\ref{app:reweight}.

\section{Zero-temperature observables}
\label{sec:zeroT}
\subsection{The residual mass}

\begin{figure}[!htbp]
 	\centering
 	\includegraphics[scale=0.65]{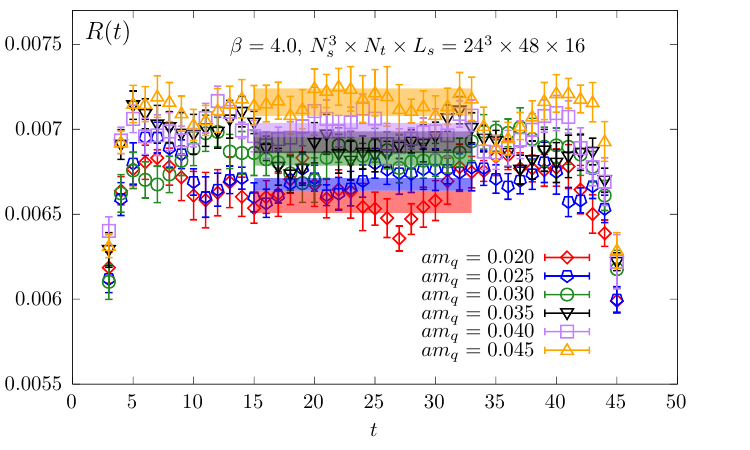}
 	\caption{The ratio $R(t)$ of the mid-point correlator to the pion correlator 	as a function of source-sink separation $t$ on  zero-temperature  $24^3\times 48\times 16$ ensembles at $\beta = 4.0$ with several quark masses. The horizontal bands indicate 
 		constant fits over $15\leq t\leq 33$, from which $am_{\rm res}(am_q)$ is determined.}
 	\label{fig:Ratio}
 \end{figure}

To quantify the residual chiral symmetry breaking 
of M\"obius domain-wall fermions with finite $L_s$, we measure 
 $\mres$ using the ratio $R(t)$ in Eq.~\eqref{eq:ratio}. 
The correlation functions are 
computed using a $Z_2$ wall source.
The ratio 
 $R(t)$ is shown in 
 \autoref{fig:Ratio} for the zero-temperature $24^3\times 48\times 16$ ensembles at $\beta=4.0$ with several different quark masses. In this plot, the horizontal bands denote fits to a constant over the range $15\le t \le 33$ with jackknife errors, from which $a\mres$ is determined for each $am_q$. 

  Although $a\mres$ should be a constant providing a universal description of residual chiral symmetry breaking effects on long-distance observables, the estimator defined in Eq.~\eqref{eq:ratio} depends on the input quark mass.

\begin{figure}[!htbp]
	\centering
	\includegraphics[scale=0.65]{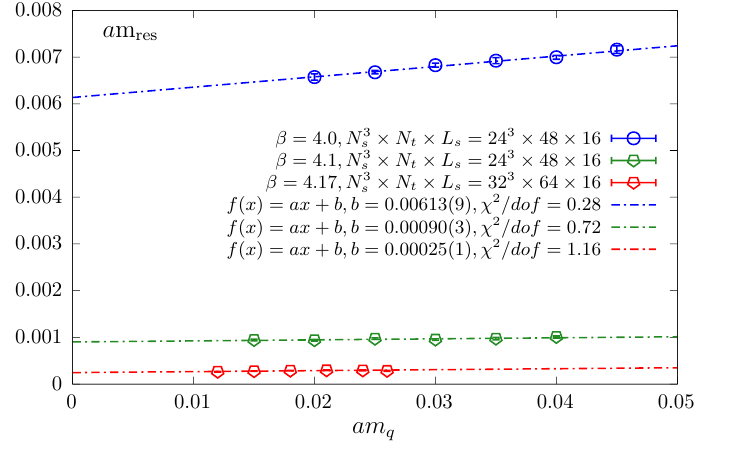}
	\caption{The residual mass as a function of the bare input quark mass for zero-temperature ensembles at $\beta=4.0$, $4.1$, and $4.17$. The dash-dotted lines 
		represent linear fits $f(x) = ax + b$, with the intercept $b$ 
		giving $am_\text{res}(0)$.} 
	\label{fig:mres_m}
\end{figure}

Figure~\ref{fig:mres_m} shows the residual mass 
as a function of bare input quark mass $a\mq$ for zero-temperature ensembles at $\beta=4.0$, 4.1, and 4.17. The dash-dotted lines represent the linear fits that describe the data well. 
This linear quark mass dependence is understood as a lattice artifact~\cite{Sharpe:2007yd, Sharpe:2006re}: 
 		$a\mres(a\mq) = a\mres(a\mq=0) \left[1+\mathcal{O}\left((a\mq) (a\Lambda)\right)\right]$. 
 		We extrapolate $a\mres$ to the zero input quark mass limit and obtain $a\mres(0) = 0.00613(9)$, 0.00090(3), and 0.00025(1) at 
$\beta=4.0$, $4.1$ and $4.17$, respectively.

\begin{figure}[!htbp]
	\centering
	\includegraphics[scale=0.65]{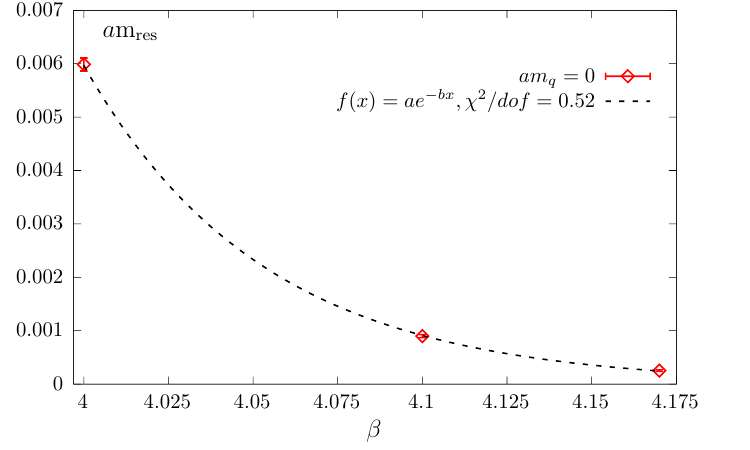}
	\caption{ The residual mass $am_\text{res}(0)$ extrapolated 
	 to zero input quark mass as a function of the gauge coupling $\beta$. The dashed line shows an exponential fit 
	 $f(\beta) = ae^{-b\beta}$, where $am_\text{res}$ decreases 
	 exponentially as $\beta$ increases.} 
	\label{fig:mres_beta}
\end{figure}

These values  strongly 
depend on $\beta$, as shown in~\autoref{fig:mres_beta}. 
 We find that $a\mres$ increases exponentially as $\beta$ decreases. This can be understood by the appearance of 
 more localized lattice dislocations 
  towards the coarser lattices~\cite{HotQCD:2012vvd}. We also verify that extrapolating to $m_q=0$ and $m_q = - \mres$ yields consistent results within errors, as shown in~\autoref{fig:sup_mres}.

\subsection{Chiral condensate}

\begin{figure*}[!htbp]
	\centering
	      \includegraphics[width=0.48\columnwidth]{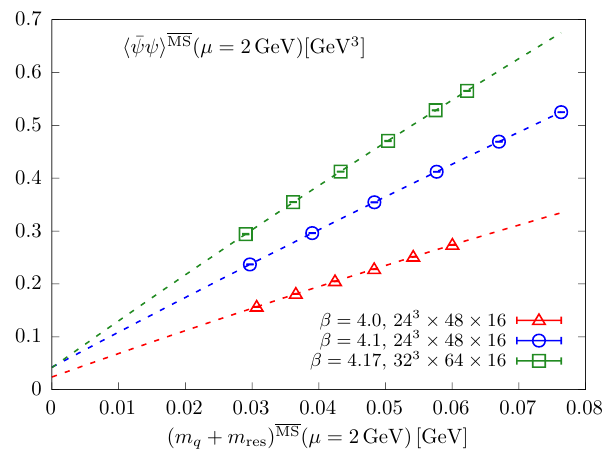}
	        \includegraphics[width=0.48\columnwidth]{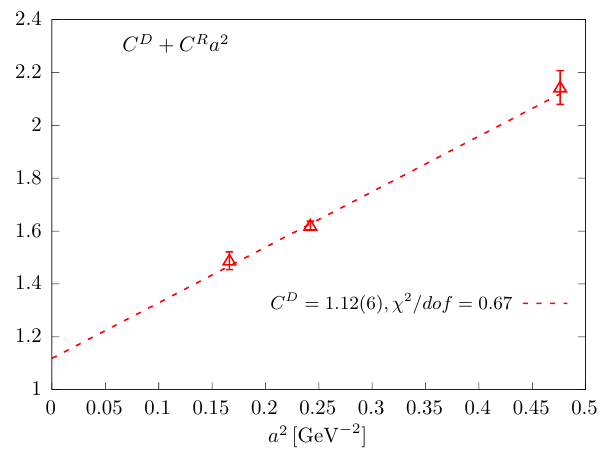}
	\caption{
	Left: The multiplicatively renormalized chiral condensate $\langle\bar\psi\psi\rangle^{\overline{\mathrm{MS}}}
	(\mu=2\,\mathrm{GeV})$ as a function of the renormalized quark mass $(m_q+m_\text{res})^{\overline{\mathrm{MS}}}(\mu=2\,\mathrm{GeV})$ on zero-temperature ensembles at $\beta=4.0$, $4.1$, and $4.17$. The ensemble sizes shown in the legend are given as $N_s^3\times N_t\times L_s$. The dashed curves show quadratic fits of the form $\barpsi(m_R,a_i)=A_i m_R^2+B_i m_R+C_i$. Right: The corresponding dimensionless linear coefficients $a_i^2B_i=C^D+C^R a_i^2$ are plotted as a function of $a_i^2$. The dashed line shows a linear extrapolation to the continuum limit, yielding $C^D=1.12(6)$.} 
	\label{fig:pbp_zeroT}
\end{figure*}

The left panel of \autoref{fig:pbp_zeroT} shows 
the chiral condensate $\barpsi$, renormalized in the $\overline{\mathrm{MS}}$ scheme at $\mu = 2\, \mathrm{GeV}$, as a function of the renormalized quark mass 
\[
m_R \equiv (m_q+m_{\mathrm{res}})^{\overline{\mathrm{MS}}}(\mu=2\,\mathrm{GeV})
= Z_m^{\overline{\mathrm{MS}}}(\mu=2\,\mathrm{GeV})\,(\mq+\mres)
\]
for $\beta=4.0, 4.1$, and $4.17$, with dashed lines showing quadratic fits. Even after multiplicative renormalization, the chiral condensate contains an additive UV-divergent contribution proportional to the quark mass, whose coefficient diverges as $1/a^2$. To separate this UV-divergent contribution from the regular mass dependence, we study the mass dependence of $\barpsi$ at several lattice spacings. For each lattice spacing $a_i$, the condensate is fitted as 
\begin{align}\label{eq:chi_condensate_fit}
\barpsi(m_R,a_i)
&=
\barpsi(0,a_i) + C^D \frac{m_R+(x-1)m_{\mathrm{res},R}}{a_i^2} + C^R m_R + A_i m_R^2 \nonumber \\
&= \underbrace{\left[
\barpsi(0,a_i) + C^D\frac{(x-1)m_{\mathrm{res},R}}{a_i^2} \right]}_{C_i} + \underbrace{\left[
\frac{C^D+C^R a_i^2}{a_i^2}
\right]}_{B_i}
m_R + A_i m_R^2 ,
\end{align}
where $m_{\mathrm{res},R}=Z_m^{\overline{\mathrm{MS}}}(\mu=2\,\mathrm{GeV})\,m_{\mathrm{res}}$. The key point of Eq.~\eqref{eq:chi_condensate_fit} is that the coefficient of the term linear in the quark mass $(B_i)$ contains two contributions: the additive UV-divergent part, $C^D/a_i^2$, and the regular part, $C^R$. Equivalently, $a_i^2B_i=C^D+C^R a_i^2$, which is the dimensionless quantity plotted in the right panel of \autoref{fig:pbp_zeroT} as a function of the squared lattice spacing $a_i^2$. Fitting $a_i^2B_i$ linearly in $a_i^2$ and extrapolating to the continuum limit, we cleanly isolate the UV-divergent coefficient, yielding $C^D=1.12(6)$.
  
\begin{figure}[!htbp]
  \centering
 \includegraphics[scale=0.65]{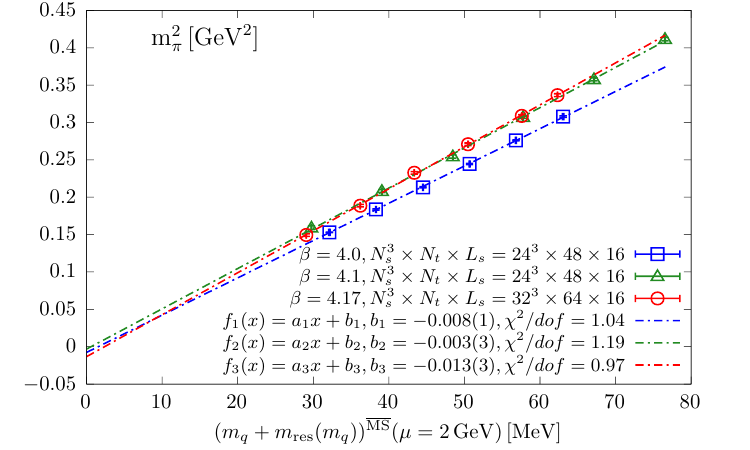}
	\caption{The pion mass squared $m_\pi^2$, 
		calculated from the pion correlator $\langle \pi^a(x)\pi^a(0)\rangle$, 
		as a function of the renormalized quark mass $(m_q+m_\text{res})^{\overline{\mathrm{MS}}}(\mu=2\,\mathrm{GeV})$ on zero-temperature ensembles at three different $\beta$ values. The dash-dotted lines represent linear fits.}
  \label{fig:mpi_mass}
\end{figure}

In \autoref{fig:mpi_mass}, we show the pion mass squared 
as a function of the renormalized 
quark mass on zero-temperature 
ensembles for three different $\beta$ values. 
The results exhibit good linearity, as indicated by the dash-dotted lines. This behavior is consistent with the leading order prediction of chiral perturbation theory, where $m_{\pi}^2 \propto m_{q} + m_{\rm{res}}$. As expected, $m_{\pi}^2$ approaches zero in the chiral limit, demonstrating the excellent chiral properties of M\"obius domain-wall fermions. However, $m_{\pi}^2$ does not reach exactly zero, which could be due to finite-volume effects or the absence 
 of chiral logarithm corrections in the linear fit ansatz.

\section{Finite-temperature observables}
\label{sec:finiteT}

\subsection{Plaquette and plaquette susceptibility}
\begin{figure}[!htbp]
  \centering
\includegraphics[scale=0.65]{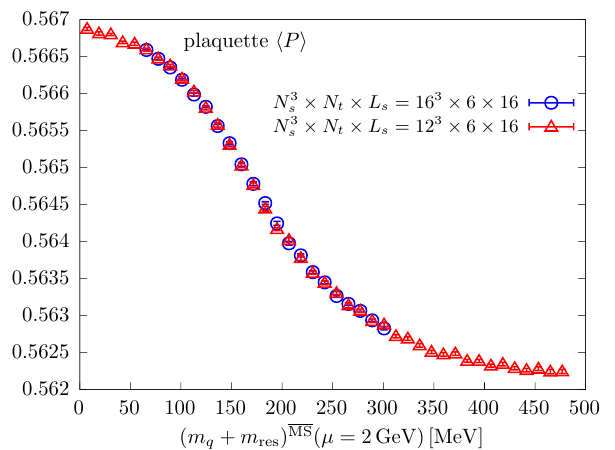}
\includegraphics[scale=0.65]{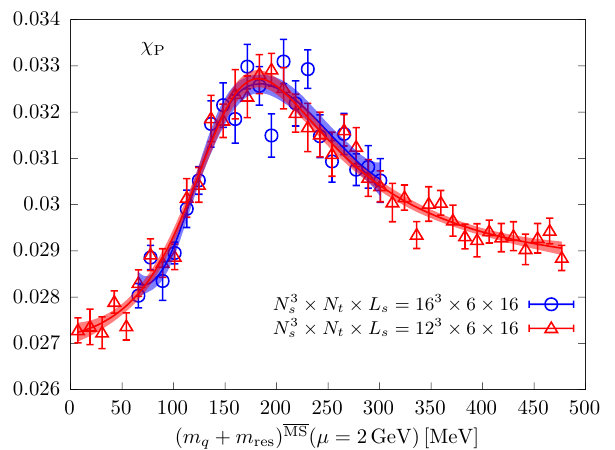}
	\caption{Left: The plaquette expectation value $\langle P \rangle$ as a function of the renormalized quark mass for $N_t=6$, corresponding to a temperature of $T = 242(4)$ MeV, with two different volumes. Right: The corresponding plaquette susceptibility $\chi_{P}$. The continuous bands in the right panel are obtained from a Pad\'e fit of order $[3/2]$. The susceptibility peaks provide estimates of 
the pseudocritical quark masses, 
$(m_q+m_{\mathrm{res}})^{\overline{\mathrm{MS}}}(\mu=2\,\mathrm{GeV})
=182(5)$ MeV and $184(10)$ MeV for the $12^3\times6\times16$ and 
$16^3\times6\times16$ ensembles, respectively, corresponding in both cases to $m_\pi\simeq0.96$~GeV.}
	\label{fig:Nt6_plaquette}
\end{figure}

We utilize the plaquette expectation value $\langle P \rangle$ and its susceptibility $\chi_{P}$ to determine the transition point. These quantities are defined as follows:
\begin{align}
 \langle P \rangle &= \frac{1}{6 N_s^3 N_t}\sum_{\vec{x},t}
\sum_{\mu < \nu} \left[1 - \frac{1}{3} \text{Re}\,\text{Tr}\, 
U_{\mu\nu}(\vec{x},t)\right],
\label{eq:plaquette}
\end{align}
\begin{align}
\chi_{P}  &= N_s^3 N_t \Bigl(  \langle P^2 \rangle - \langle P \rangle^2 \Bigr).
\label{eq:plaquette_sus}
\end{align}
  Figure~\ref{fig:Nt6_plaquette} shows the plaquette and its susceptibility plotted as a function of the renormalized quark mass at $N_t=6$ and $\beta=4.0$, corresponding to the temperature of 242(4) MeV, for two different volumes. The region of rapid change in the plaquette aligns with a peak in the susceptibility. As shown in the right panel, the peaks of $\chi_{P}$ are located at 182(5) MeV and 184(10) MeV on the $12^3\times 6 \times 16$ and $16^3\times 6 \times 16$ lattices, respectively, providing estimates of the pseudocritical quark masses, corresponding in both cases to
  $m_\pi\simeq0.96$~GeV. The continuous bands are obtained from a Pad\'e fit of order $[3/2]$. The negligible volume dependence of the $\chi_{P}$ peak height suggests a crossover transition.

\subsection{Chiral condensate}
\begin{figure}[!htbp]
  \centering
			\includegraphics[scale=0.65]{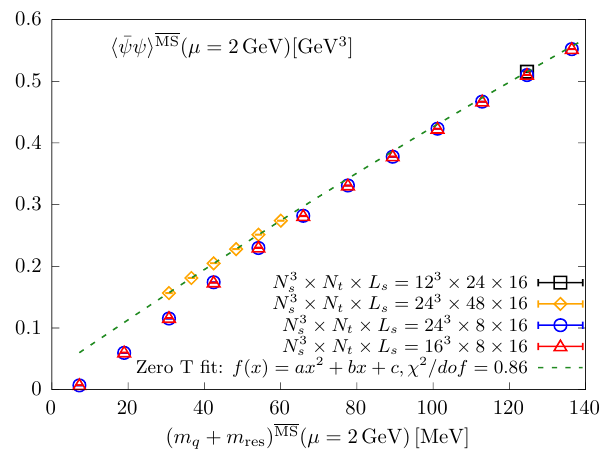}
		\includegraphics[scale=0.65]{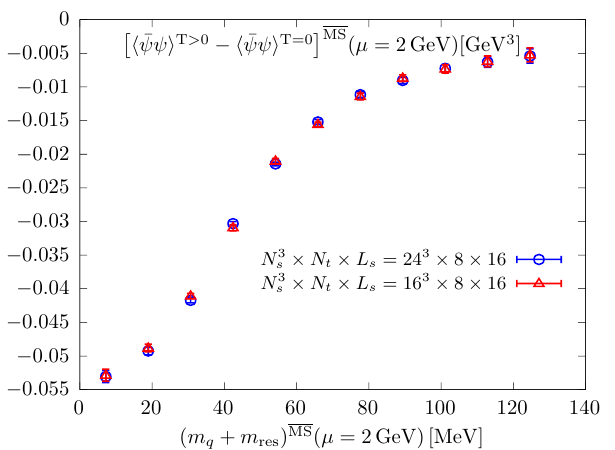}
	\caption{Left: The multiplicatively renormalized chiral condensate on finite-temperature lattices ($16^3\times8$, $24^3\times8$) and zero-temperature lattices ($12^3\times24$, $24^3\times48$) for $\beta=4.0$ is shown as a function of the renormalized quark mass. The dashed curve is a quadratic fit to the zero-temperature data, including
		the $12^3\times24$ point at $(m_q+m_{\rm res})^{\overline{\rm MS}}(\mu=2\,{\rm GeV})=125$~MeV,
		which constrains the fit at large quark mass; this is the same fit used for
		the subtraction shown in the right panel. Right:  
		The subtracted chiral condensate
	 $[\langle\bar\psi\psi\rangle^{T>0} - 
		\langle\bar\psi\psi\rangle^{T=0}]^{\overline{\mathrm{MS}}}
		(\mu=2\,\mathrm{GeV})$ as a function of the renormalized quark mass for lattices of size $16^3\times8 \times 16$ and $24^3\times8 \times 16$.}
	\label{fig:Nt8_pbp}
\end{figure}

The left panel of~\autoref{fig:Nt8_pbp} depicts the multiplicatively renormalized chiral condensate  $\langle\bar{\psi}\psi\rangle^{\overline{\mathrm {MS}}}(\mu=2\,\mathrm{GeV})$, 
without subtracting the additive 
divergence, as a function of the renormalized quark mass. 
The data are obtained on 
zero-temperature lattices and finite-temperature lattices with $N_{t}=8$, both at 
$\beta = 4.0$, where the finite-temperature lattices span two different volumes with 
aspect ratios $N_{s}/N_{t}=$ $2$ and $3$. 
We observe that $\langle\bar{\psi}\psi\rangle^{\overline{\mathrm {MS}}}(\mu=2\,\mathrm{GeV})$ on the $N_{t}=8$ lattices 
vanishes at a positive value of quark mass. This is because $x \neq 1$: the UV-divergent term $C^D(m_q+x\,m_{\mathrm{res}})a^{-2}$ 
vanishes at $m_q=-x\,m_{\mathrm{res}}$, rather than at the chiral limit 
$m_q=-m_{\mathrm{res}}$. Therefore, at $m_q+m_{\mathrm{res}}=0$, the unsubtracted chiral condensate still contains a residual contribution $C^D(x-1)m_{\mathrm{res}}/a^{2}$. 
As will be shown below, $x < 1$, so this residual term is negative 
and shifts the zero of the unsubtracted chiral condensate to a positive renormalized quark mass.
 
The UV divergence can be eliminated by subtracting the zero-temperature chiral condensate from the finite-temperature data at the same quark mass. However, this requires zero-temperature lattices at $\beta=4.0$ with the same quark masses as the finite-temperature ensembles, and some of these masses are not available in our zero-temperature simulations. To address this, we perform a quadratic extrapolation of the zero-temperature chiral condensate, shown as the dashed line in the left panel of~\autoref{fig:Nt8_pbp}. This procedure 
 yields the additively and multiplicatively renormalized chiral condensate
 $[\barpsi^{\mathrm{T}>0} - \barpsi^{\mathrm{T}=0}]^{\overline{\mathrm {MS}}}(\mu=2\,\mathrm{GeV})$, 
 presented in the right panel of~\autoref{fig:Nt8_pbp}.

We observe volume independence of the chiral condensate for lattices of size $24^3\times8$  and $16^3\times8$ in both panels, since the UV divergence term is volume independent.
The right panel shows a smooth, monotonic increase of the subtracted condensate toward zero as the renormalized quark mass increases, with no sign of a discontinuity. 
This indicates that the transition is an analytic crossover. A precise determination of the transition quark mass from the chiral condensate alone is difficult due to its gradual variation; we therefore defer to the chiral susceptibility, which exhibits a peak that defines the crossover location.

\begin{figure}[!htbp]
  \centering
\includegraphics[scale=0.65]{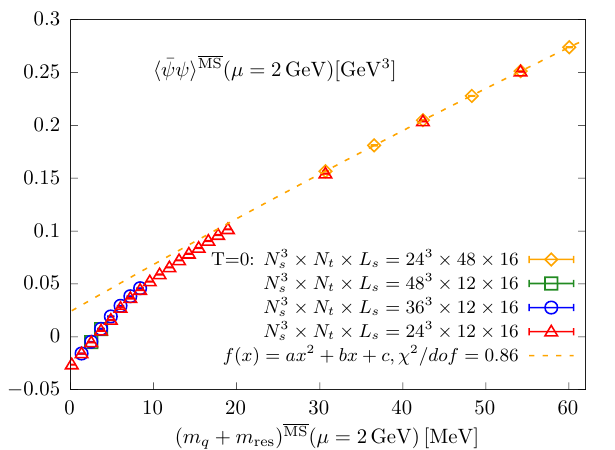}
\includegraphics[scale=0.65]{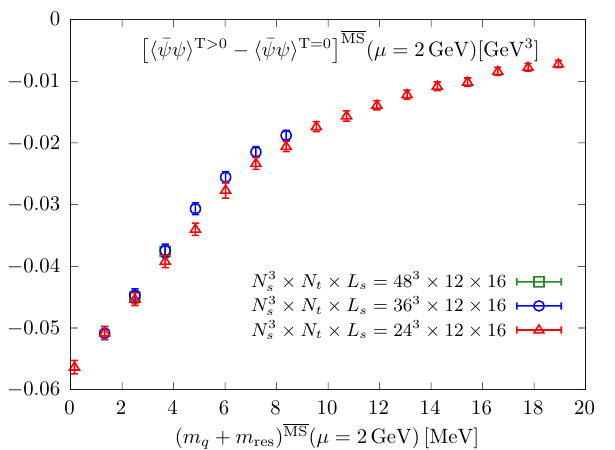}
			\caption{Same as Fig.~7, but for finite-temperature ensembles with $N_t = 12$ (volumes $24^3\times12$, $36^3\times12$, and $48^3\times12$) at $\beta = 4.0$. Zero-temperature data are from the $24^3\times48 \times 16$ ensembles. The dashed curve in the left panel represents 
				the extrapolation of the zero-temperature chiral condensate. 
				Right: The subtracted chiral condensate 
				$[\langle\bar\psi\psi\rangle^{T>0} - 
				\langle\bar\psi\psi\rangle^{T=0}]^{\overline{\mathrm{MS}}}
				(\mu=2\,\mathrm{GeV})$ as a function of the renormalized 
				quark mass for $N_t=12$ ensembles.}
	\label{fig:Nt12_pbp}
\end{figure}

In \autoref{fig:Nt12_pbp}, we present the same quantities as in~\autoref{fig:Nt8_pbp}, but for the $N_{t}=12$ ensembles. In the left panel, the unsubtracted chiral condensate becomes negative near the chiral limit. This negative value arises from the residual additive UV-divergent contribution, $C^D (x-1)\mres a^{-2}$. In the right panel, the zero-temperature subtraction removes this additive UV-divergent contribution. The subtracted condensate remains negative near the chiral limit because the finite-temperature condensate is smaller than the zero-temperature condensate. At larger quark masses, the chiral condensate becomes less sensitive to finite-temperature effects, so the finite- and zero-temperature condensates become closer and the subtracted condensate moves toward zero. We observe visible finite-volume effects around $(m_q + \mres)^{\overline{\mathrm {MS}}}(\mu = 2\,\mathrm{GeV}) \sim 4\text{--}9$ MeV. It is difficult to determine whether this is a crossover or a true phase transition, and the inflection point is also hard to locate just from the chiral condensate itself.  To address this, we calculate the disconnected chiral susceptibility and total chiral susceptibility.

To study the $N_f=3$ QCD phase transition in the chiral region, small quark mass ($m = m_q + \mres$) simulations are needed. The simplest way to achieve this is to use a negative input quark mass $m_q$; however, this carries the risk of inducing singularities in the Dirac operator during the dynamical evolution when the magnitude of the negative input quark mass exceeds the corresponding residual mass~\cite{HotQCD:2012vvd}. In our simulations, the magnitude of
the most negative input quark mass
 we used was roughly comparable to the 
residual mass, and we did not observe any singularity issues. Another way of achieving the target small value of the total quark mass is by increasing the value of $L_s$, since $\mres$ decreases as $L_s$ is increased.

\begin{figure}[!htbp]
	\centering
	\includegraphics[scale=0.65]{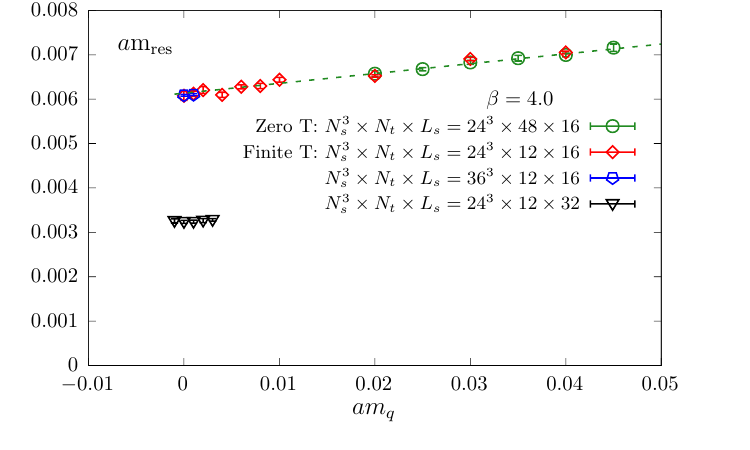}
	\caption{$a\mres$ as a function of the bare input quark mass for the zero-temperature lattices with $L_s=16$ and the finite-temperature lattices with $L_s=16$ and $L_s=32$. }
	\label{fig:mres_difLs}
\end{figure}

Since simulations with negative input quark masses are uncommon for M\"{o}bius domain-wall fermions, we performed an additional cross-check by generating $24^3\times 12 \times 32$ ensembles to assess residual chiral symmetry breaking effects and the reliability of the large negative input quark masses used for the $N_t=12,\, L_s=16$ ensembles. The input quark masses for these $L_s=32$ ensembles were chosen to target approximately the same total quark mass as the corresponding $L_s=16$ ensembles. We measured $\mres$ using the ratio of spatially separated mid-point correlation functions to pion correlation functions, as opposed to the temporally separated ones used in Eq.~\eqref{eq:ratio}, since the temporal extent is rather short at finite temperature. As shown in \autoref{fig:mres_difLs}, $\mres$ for $L_s=32$ is reduced by a factor of two relative to $L_s=16$, consistent with the expected $1/L_s$ dependence. At strong gauge 
coupling, this behavior is expected because $m_{\rm{res}}$ receives significant contributions from near-zero eigenmodes, which yield a power-law dependence of $1/L_s$ due to the presence of gauge field dislocations~\cite{RBC:2008cmd}. 
\autoref{fig:mres_difLs} also demonstrates that the values of $\mres$ for $L_s=16$, computed at a fixed $\beta=4.0$, are consistent between zero-temperature and finite-temperature ensembles, showing no volume dependence. 
 
\begin{figure}[!htbp]
	\centering
	%
		\includegraphics[scale=0.65]{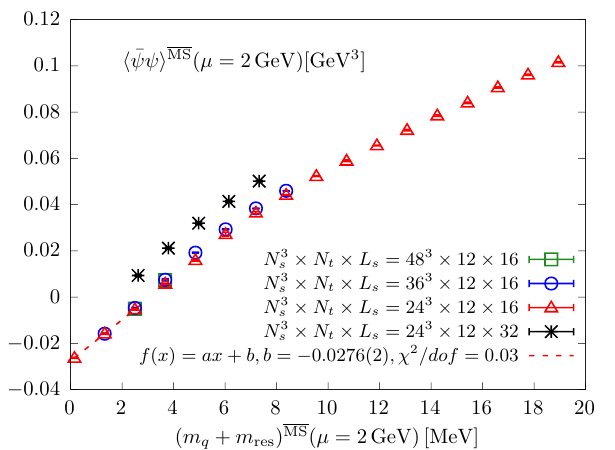}
	\includegraphics[scale=0.65]{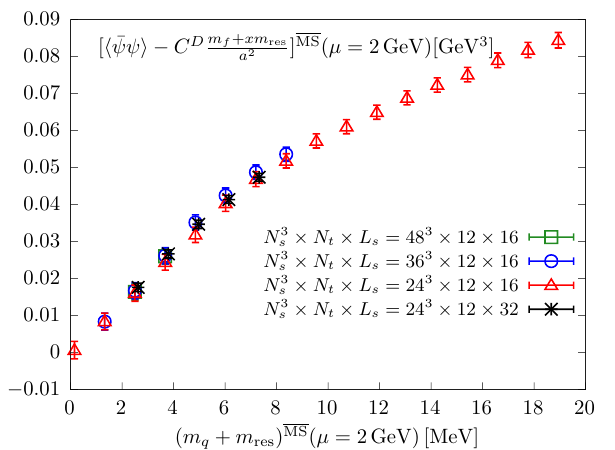}
	\caption{The multiplicatively renormalized chiral condensate (left) and the additively and multiplicatively renormalized chiral condensate (right) as functions of the renormalized quark mass for $N_t=12$ lattices with $L_s=16$ and $L_s=32$.}
	\label{fig:pbp_Nt12}
\end{figure}

More importantly, after explicitly subtracting the additive UV-divergent contribution $C^D(m_q+x\mres)/a^2$ from the chiral condensate, the renormalized subtracted condensates from the $L_s=16$ and $L_s=32$ ensembles agree at the same total quark mass, as shown in the right panel of~\autoref{fig:pbp_Nt12}. This agreement, together with the consistency of the chiral susceptibilities, confirms the reliability of the $N_t=12$, $L_s=16$ simulations with negative input quark masses.

From the left panel of~\autoref{fig:pbp_Nt12}, we observe that close to the chiral limit, the chiral condensate is negative for the $24^3 \times 12 \times 16$ ensemble. This is due to the lack of exact chiral symmetry, as shown by the expression: $\lim _{m_q + \mres \to 0} \barpsi|_{\mathrm{DWF}} \sim C^D \frac{(x-1 )\mres}{a^2} + \barpsi|_{\mathrm{cont}} + \cdots$. The negativity of the chiral condensate implies $x<1$. We also see that increasing $L_s$ from 16 to 32 while keeping the total quark masses roughly the same causes an increase in the chiral condensate. This 
behavior is straightforward to understand by rewriting
Eq.~\eqref{eq:DWF_pbp} as $\barpsi|_{\mathrm{DWF}} \sim C^D \big(\frac{(x-1 )\mres}{a^2} +  \frac{m_q +\mres}{a^2}\big) + \barpsi|_{\mathrm{cont}}+ \cdots$. Since $\mres$ is smaller for $L_s=32$ than $L_s=16$ and $x<1$, for similar total quark masses, the chiral condensate result for $L_s=32$ is always larger than for $L_s=16$. We expect that with $L_s=32$, the residual chiral symmetry breaking effect will become smaller, and this can be seen if we do the extrapolation to the chiral limit. 

The unwanted UV divergence component, $C^D \frac{m_q + x\mres}{a^2}$, can be removed explicitly without contaminating the regular part contribution if the value of $x$ is known. The value of $x$ can be estimated using the finite-temperature chiral condensate results on $24^3 \times 12 \times 16$ lattices, 
evaluated at quark masses close to the chiral limit.
 In the chiral limit, the chiral condensate takes the form:  $\lim_{m_q + \mres \to 0}\barpsi|_{\mathrm{DWF}} \sim  \barpsi|_{\mathrm{cont}} + C^D \frac{(x-1)\mres}{a^2}$. 
We believe that at this temperature, $T = 121(2)$ MeV, chiral symmetry is already restored in the chiral limit, according to the estimate $T_c = 98^{+3}_{-6}\, \text{MeV}$ for $N_t=8$~\cite{Dini:2021hug}, and therefore assume $\barpsi|_{\mathrm{cont}}=0$. Using the linear fit result in the chiral limit---indicated by the intercept of the dashed line, which equals $C^D \frac{(x-1)\mres}{a^2}$---together with the known value of $\mres$ for $L_s=16$ and $C^D$, we obtain $x=-0.6(1)$.
With the knowledge of $\mres$ for $L_s=32$, and assuming that $x$ is independent of $L_s$ at fixed gauge coupling, we can now effectively eliminate the additive UV divergence term for both the $L_s=16$ and $L_s=32$ lattices. 
The right panel of~\autoref{fig:pbp_Nt12} displays the renormalized subtracted chiral condensate results. 

We observe that the results from the $24^3\times 12\times16$ and $24^3\times 12\times 32$ lattices exhibit consistency, with some minor discrepancies likely attributed to slight differences in total quark mass. This observation implies that the UV divergence term, which includes the residual chiral symmetry breaking effect, has been successfully subtracted from the chiral condensate.

\begin{figure}[!htbp]
	\centering
	\includegraphics[scale=0.85]{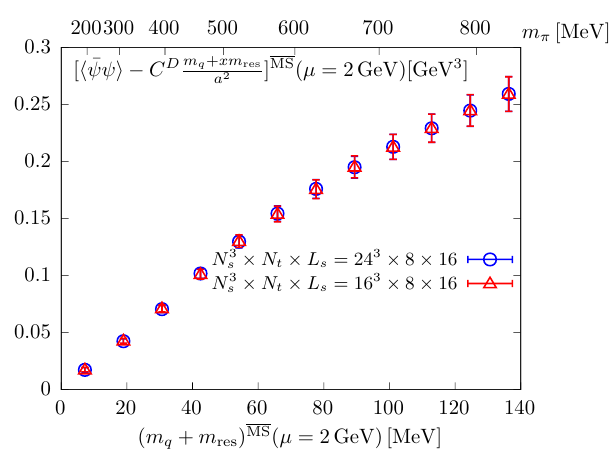}
	\caption{ The renormalized subtracted chiral condensate for $N_t=8$ lattices with two different volumes. The upper axis indicates the corresponding
		pseudoscalar mass; these are leading-order estimates meant for orientation
		only, see the beginning of Appendix~\ref{app:params}.} 
	\label{fig:Nt8_sub_pbp}
\end{figure}

\autoref{fig:Nt8_sub_pbp} shows the renormalized subtracted chiral condensate  for $N_t=8$ lattices. 
A simple linear extrapolation would 
show that the renormalized subtracted chiral condensate vanishes in the chiral limit. This is expected since the current temperature is safely above 
the critical temperature.

\subsection{Chiral susceptibility}

On our finite-temperature lattices, we calculate the chiral condensate using 10  stochastic noise vectors for each gauge configuration. This allows us to estimate the fluctuations of the chiral condensate---in other words, the disconnected chiral susceptibility---in an unbiased way.
\begin{figure}[!htbp]
	\centering
	\includegraphics[scale=0.65]{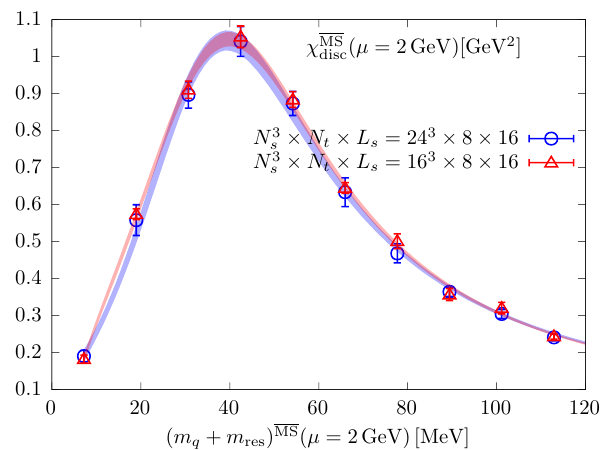}
		\includegraphics[scale=0.65]{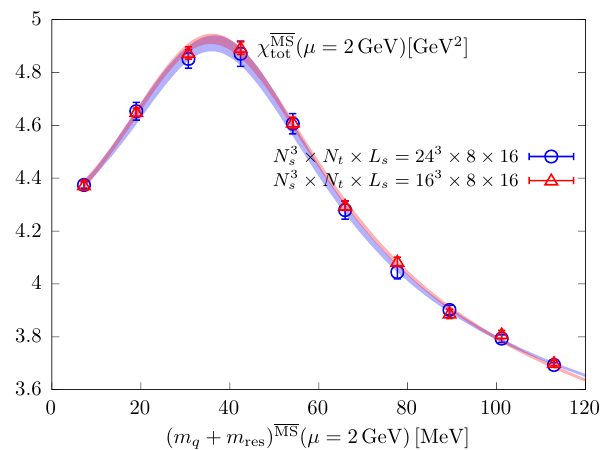}
	\caption{The renormalized disconnected chiral susceptibility (Left) and the multiplicatively renormalized total chiral susceptibility (Right) as functions of the renormalized quark mass for $N_t = 8$ lattices, corresponding to $T=181(3)$ MeV, with two different volumes, along with their Pad\'e fits.
    The susceptibility peaks give pseudocritical quark masses of
$(m_q+m_{\rm res})^{\overline{\rm MS}}(\mu=2\,{\rm GeV}) = 39.1(9)$--$40.2(6)$~MeV
(disconnected) and $36(1)$--$36.1(7)$~MeV (total), for the
$24^3\times8\times16$ and $16^3\times8\times16$ ensembles, respectively, i.e.\ $m_\pi\simeq445$--$451$~MeV (disconnected) and
$\simeq427$~MeV (total).}
	\label{fig:chi_disc_Nt8}
\end{figure}

The left and right panels of~\autoref{fig:chi_disc_Nt8} show the renormalized disconnected chiral susceptibility $\chi_{\mathrm{disc}}^{\overline{\mathrm {MS}}}(\mu=2\,\mathrm{GeV})$ and the multiplicatively renormalized total chiral susceptibility $\chi_{\mathrm{tot}}^{\overline{\mathrm {MS}}}(\mu=2\,\mathrm{GeV})$ as functions of the renormalized quark mass for lattices of size $N_{s}^3 \times 8$ with $N_{s} = 16$ and 24, respectively. 
The total susceptibility 
still suffers from the additive divergence. However, since we are using a fixed lattice spacing, the additive divergence is identical across different volumes and therefore does not affect our analysis of the critical behavior. 
 The bands represent Pad\'e
 fits. We observe that both chiral susceptibilities exhibit no significant volume dependence,
which indicates an analytic crossover. 
The disconnected susceptibility exhibits a pronounced peak at 
a quark mass of 40.2(6) MeV for $16^3\times 8$ lattices and 39.1(9) MeV for $24^3\times 8$ lattices, while the total susceptibility develops a peak at 36.1(7) MeV for $16^3\times 8$ lattices and 36(1) MeV for $24^3\times 8$ lattices. 
 Since the transition region is broad for 
 a crossover, the pseudocritical point determined from different observables can differ.  

\subsection{Quark mass reweighting}

Generating gauge configurations with M\"obius domain-wall fermions is computationally expensive for large volumes and light quark masses. Due to limited computational resources, we simulated only two mass points for $48^3 \times 12 \times 16$ lattices, positioned near the transition region. To precisely locate the transition point, additional mass points are required. Instead of generating new gauge configurations, we employ reweighting of the sea quark mass, which is significantly cheaper. This technique reuses an ensemble generated at one quark mass to obtain predictions at a different mass. The accessible quark mass range via reweighting is constrained by the overlap in configuration space between the generated and target ensembles; if the overlap is too small, the reweighting factor exhibits large fluctuations.

Quark mass reweighting requires computing a
reweighting factor for each gauge configuration. In this work, we use three degenerate flavors of M\"obius domain-wall fermions. For an ensemble of gauge configurations generated at sea quark mass $m_1$, we can reweight it to an ensemble with sea quark mass $m_2$ using a reweighting factor $w(U; m_1, m_2)$, defined as the ratio of fermion determinants~\cite{Hasenfratz:2008fg,Liu:2012gm}:
  \begin{align}\label{eq:reweighting_factor}
w(U; m_1, m_2) = \Bigg(\frac{\det [D_{\text{MDWF}}^{\dagger}(U, m_2) D_{\text{MDWF}}(U, m_2)]} {\det [D_{\text{MDWF}}^{\dagger}(U, m_1) D_{\text{MDWF}}(U, m_1)]}\Bigg)^{3/2} = \Big(\frac{1}{\det A}\Big)^{3/2}\,, 
\end{align}
where $D_{\text{MDWF}}(U, m_1)$ is the M\"obius domain-wall fermion Dirac operator with fermion mass $m_1$, and the matrix $A$ is defined as $A= D_{\text{MDWF}}^\dagger(m_1) {D_{\text{MDWF}}^\dagger(m_2)}^{-1} D_{\text{MDWF}}(m_2)^{-1} D_{\text{MDWF}}(m_1)$. Since an exact computation of the reweighting factor is prohibitively expensive, we employ stochastic estimation with Gaussian noise vectors:
\begin{align}\label{eq:stochastic_reweighting_factor}
	w(U; m_1, m_2)  
	=\Bigg( \frac{\int \mathcal{D}\xi^{\dagger} \mathcal{D}\xi \,e^{-\xi^\dagger\xi} \, e^{-\xi^\dagger (A -I) \xi} }{\int \mathcal{D}\xi^{\dagger} \mathcal{D}\xi \,e^{-\xi^\dagger\xi} } \Bigg)^{3/2} = {\langle  e^{-\xi^\dagger (A -I) \xi}       \rangle_{\xi}}^{3/2}\,.
\end{align}
The expectation value of an observable $O$ at the target sea quark mass $m_2$ is then obtained by computing the weighted average over the $N$ configurations generated at $m_1$:
\begin{align}\label{reweighted_ob}
\langle O \rangle_2 =\frac{ \langle O \,w(m_1, m_2) \rangle_1}{\langle w(m_1, m_2) \rangle_1} \approx \frac{\sum_{i=1}^{N}w(U_i; m_1, m_2)O(U_i)}{\sum_{i=1}^{N}w(U_i; m_1, m_2)}\,.
\end{align}
Having calculated the
reweighting factors 
configuration by configuration, we can 
obtain the reweighted physical quantity of interest. 
We performed simulations at two bare input quark masses, $am_q=-0.003$ and $-0.004$, on the $48^3\times 12 \times 16$ lattices. 
Since these two points are 
not enough to locate the transition mass point, 
we resorted to sea quark mass reweighting. 
We have seven mass points reweighted from $am_q=-0.003$ and three mass points reweighted from $am_q=-0.004$ in order to cover the transition range. 

\begin{figure}[!htbp]
	\centering
		\includegraphics[scale=0.65]{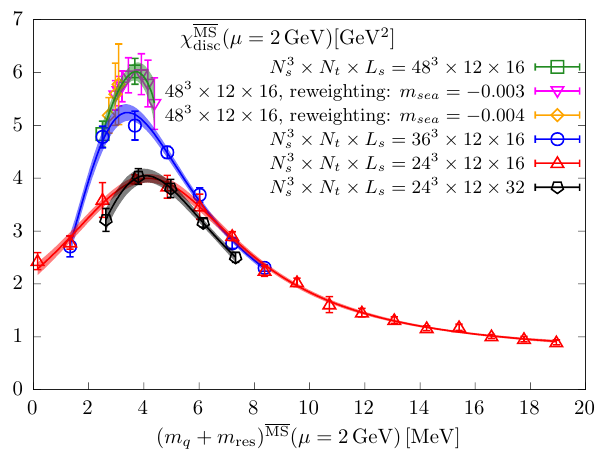}
		\includegraphics[scale=0.65]{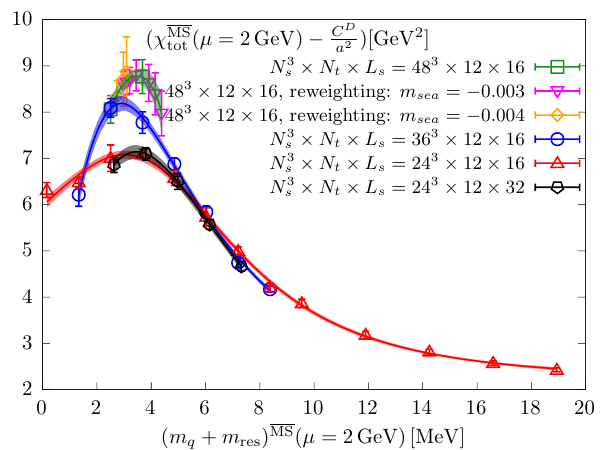}
	\caption{The renormalized disconnected chiral susceptibility (Left) and total susceptibility (Right) as functions of the renormalized quark mass for $N_t=12$ lattices with $L_s=16$ and $L_s=32$. The bands are from Pad\'e fits.}
	\label{fig:chi_disc_Nt12}
\end{figure}

In the left panel of~\autoref{fig:chi_disc_Nt12}, $\chi_{\mathrm{disc}}^{\overline{\mathrm {MS}}}(\mu=2\,\mathrm{GeV})$ shows a sizeable finite-volume effect for $N_{s}^3 \times 12 \times 16$ lattices with $N_{s} = 24, 36, 48$ 
over a broad range of $(m_q + \mres)^{\overline{\mathrm {MS}}}(\mu=2\,\mathrm{GeV}) \sim 2-6$ MeV, which is near the transition range. In the right panel of~\autoref{fig:chi_disc_Nt12}, we observe a slightly smaller finite-volume effect for  $(\chi_{\mathrm{tot}}^{\overline{\mathrm {MS}}}(\mu=2\,\mathrm{GeV}) - C^D a^{-2})[\rm{GeV}^2]$ at quark masses around $(m_q + \mres)^{\overline{\mathrm {MS}}}(\mu=2\,\mathrm{GeV}) \sim 2-5$ MeV for the same lattices. 
Here, the large, volume-independent UV divergence term $C^D/a^2$ is subtracted to prevent its constant background from obscuring
the finite-size effects. The transition mass points determined from 
the pronounced peaks of 
$\chi_{\mathrm{disc}}^{\overline{\mathrm {MS}}}(\mu=2\,\mathrm{GeV})$ and 
$(\chi_{\mathrm{tot}}^{\overline{\mathrm {MS}}}(\mu=2\,\mathrm{GeV}) - \frac{C^D}{a^2})[\rm{GeV}^2]$ for our largest volume ($48^3 \times 12 \times 16$) 
are 3.7(2) MeV and 3.5(3) MeV, respectively, in the $\overline{\mathrm{MS}}$ scheme at a scale of $\mu=2$ GeV, 
corresponding to
$m_\pi\simeq137$ and $133$~MeV. At a genuine chiral phase transition in the thermodynamic limit, these two determinations are expected to coincide. 

\begin{figure}[!htbp]
	\centering
					\includegraphics[scale=0.45]{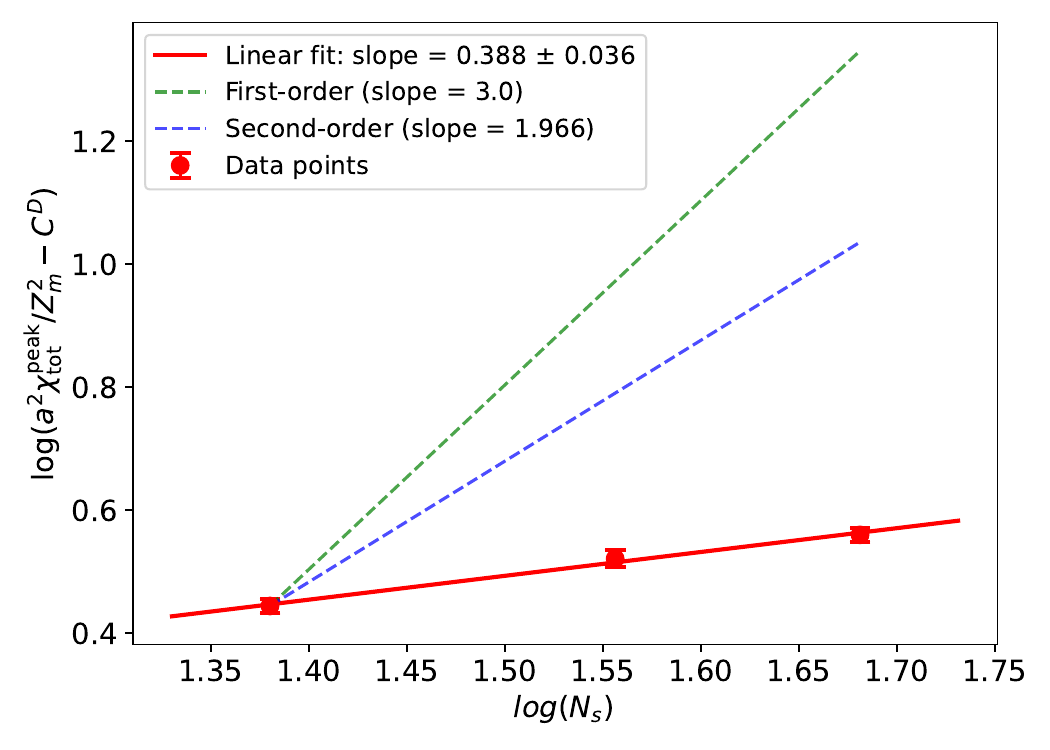}~
			\includegraphics[scale=0.316]{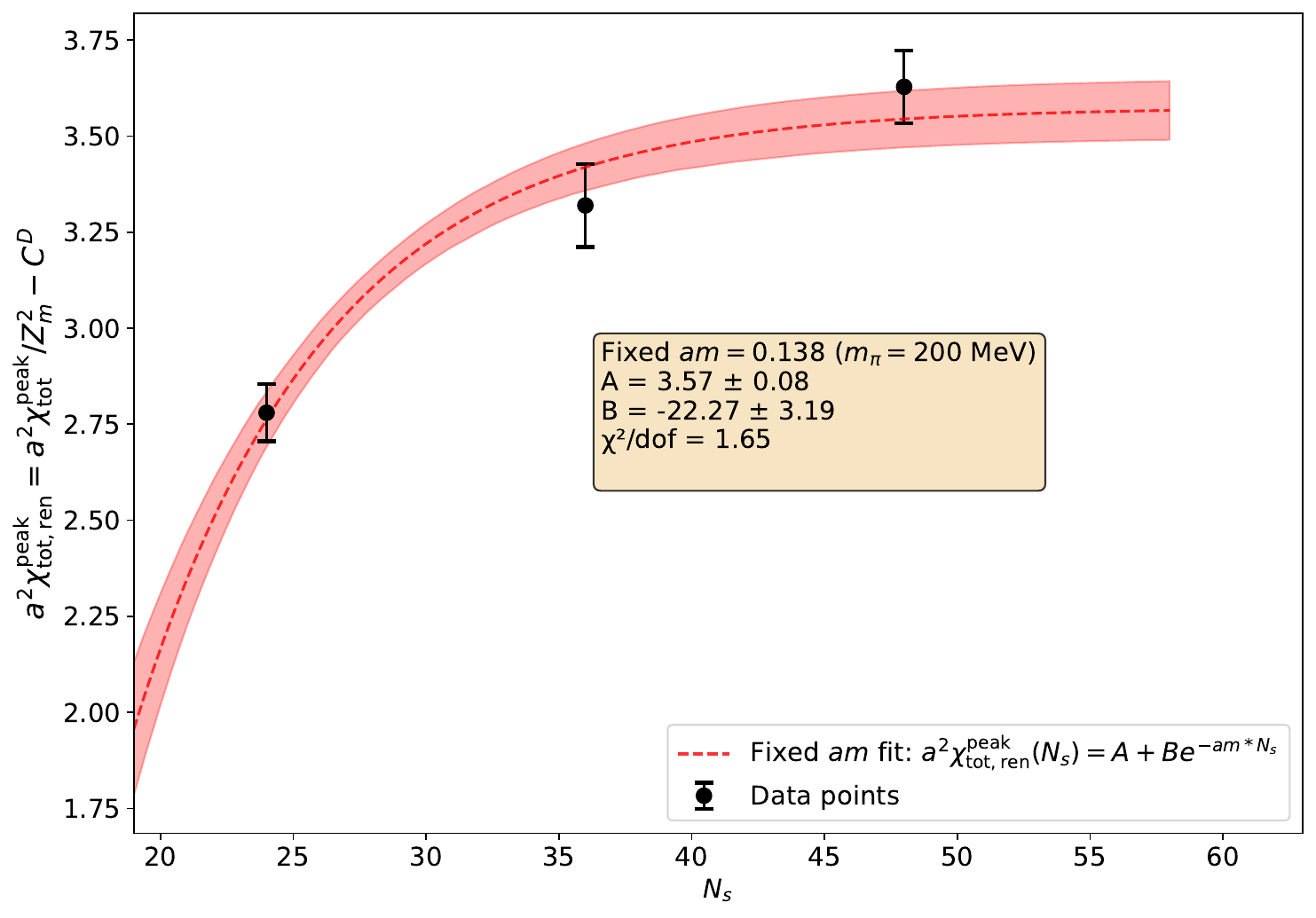}
	\caption{Left: logarithm of the peak height of the total chiral susceptibility as a function of the logarithm of the spatial extent $N_s$. Right: The exponential fit of the total chiral susceptibility as a function of $N_s$. }
	\label{fig:rescale_chi_sus_Nt12}
\end{figure}

For a crossover transition, the peak height of chiral susceptibilities should be volume independent for large volumes. However, we observe that 
the peak height continues to increase up to our largest volume, $48^3 \times 12 \times 16$.
According to the finite-size scaling analysis, the peak height of the renormalized total susceptibility is expected to be proportional to $N_s^3$ for a first-order phase transition and $N_s^{1.966}$ for a $Z(2)$ second-order phase transition.
However, as shown in the left plot of~\autoref{fig:rescale_chi_sus_Nt12}, the change in peak height 
for the total chiral susceptibility is not as large as anticipated from either a first-order or $Z(2)$ second-order phase transition. The observed $N_s^{0.39(4)}$ growth remains puzzling. Instead, we also consider a finite-volume effect that is exponentially suppressed as $e^{-amN_s}$ at large volumes. The exponent must be controlled by the lightest degree of freedom in the system. Given that we only have three different volumes, we fix the 
mass in the exponent to the measured pion screening mass near the transition region, $m_{\pi} \approx 200\,\text{MeV}$, as shown in~\autoref{fig:screening_mass},   and apply an exponential fit.  
This fit describes the data reasonably well,
as shown in the right plot of~\autoref{fig:rescale_chi_sus_Nt12}. An exponential volume dependence of the chiral susceptibility for a crossover has also been observed in Ref.~\cite{RubenKara:2023qig} using stout staggered fermions; however, in their study, the peak height decreases to reach a plateau at large volumes, whereas our results show an increase towards saturation. Taken together, the lack of expected power-law growth and the exponential saturation of the peak height most likely indicate an analytic crossover rather than a true phase transition.

Unlike the chiral condensate, the disconnected and total chiral susceptibilities exhibit a clear dependence on the total quark mass. As shown in \autoref{fig:chi_disc_Nt12}, at approximately equal total quark masses, the disconnected and total chiral susceptibilities obtained with $L_s=32$ closely agree with those obtained with $L_s=16$ on the $24^3\times 12$ lattices, despite the substantial difference in the relative magnitudes of the residual mass and the input quark mass. This behavior is expected because, upon differentiating the chiral condensate with respect to the input quark mass, both the additive UV-divergent term and the regular term linear in the quark mass become mass-independent constants, whereas the higher-order terms retain their dependence on the total quark mass. We also obtain consistent pseudocritical quark masses on the $24^3\times 12\times 16$ and $24^3\times 12\times 32$ lattices. This agreement provides an encouraging consistency check on the use of relatively large negative input quark masses for the $L_s=16$ ensembles.

Table~\ref{tab:summary} summarizes the pseudocritical quark masses obtained from the susceptibility peaks discussed in Sec.~\ref{sec:finiteT}, using the largest available
volume at each temperature.

\begin{table}[htbp]
	\centering
	\setlength{\tabcolsep}{10pt}
	\renewcommand{\arraystretch}{1.2}
	\caption{Pseudocritical quark mass $(m_q+m_{\rm res})^{\overline{\rm MS}}$ at
		$\mu=2\,{\rm GeV}$, obtained from the susceptibility peaks, together with the
		observable used and the corresponding pseudoscalar mass.}
	\label{tab:summary}
	\begin{tabular}{c c c c c c}
		\hline\hline
		$N_t$ & $T$ (MeV) & Volume & Observable &
		$(m_q+m_{\rm res})^{\overline{\rm MS}}$ (MeV) & $m_\pi$ (MeV) \\
		\hline
		6  & 242(4) & $16^3\times 6\times 16$  & $\chi_P$ & 184(10) & $\simeq960$ \\
		8  & 181(3) & $24^3\times 8\times 16$  & $\chi_{\rm disc}^{\overline{\rm MS}},\,\chi_{\rm tot}^{\overline{\rm MS}}$ & 39.1(9),\, 36(1) & $\simeq445,\,427$ \\
		12 & 121(2) & $48^3\times 12\times 16$ & $\chi_{\rm disc}^{\overline{\rm MS}},\,\chi_{\rm tot}^{\overline{\rm MS}}$ & 3.7(2),\, 3.5(3) & $\simeq137,\,133$ \\
		\hline\hline
	\end{tabular}
\end{table}

\subsection{Binder cumulant}

\begin{figure}[!htbp]
	\centering
	\includegraphics[scale=0.9]{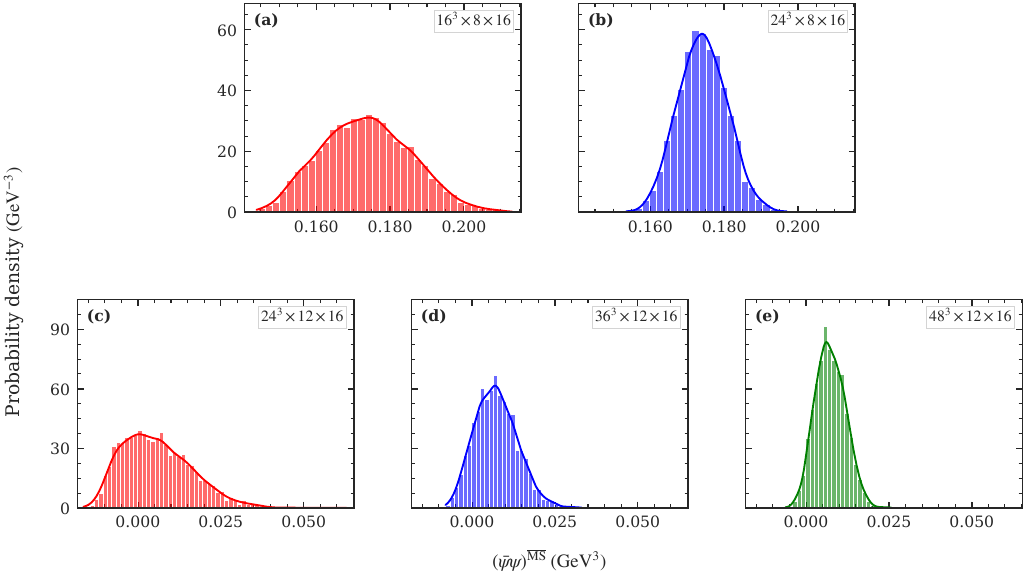}
\caption{Probability-density histograms of the chiral condensate,
multiplicatively renormalized in the $\overline{\mathrm{MS}}$ scheme at
$\mu=2\,\mathrm{GeV}$, near the transition-mass points. Panels (a) and (b)
show the $N_t=8$ ensembles at
$(m_q+m_{\mathrm{res}})^{\overline{\mathrm{MS}}}
(\mu=2\,\mathrm{GeV})\simeq42\,\mathrm{MeV}$, while panels (c)--(e)
show the $N_t=12$ ensembles at
$(m_q+m_{\mathrm{res}})^{\overline{\mathrm{MS}}}
(\mu=2\,\mathrm{GeV})\simeq3.6\,\mathrm{MeV}$.
The lattice dimensions are indicated in each panel. Within each $N_t$
group, common bin edges and common axis ranges are used for all spatial
volumes. Each histogram is normalized to unit area. The solid curves are
kernel-density estimates shown as guides to the eye.}	\label{fig:histogram_pbp_Nt12}
\end{figure}

Examining the distribution of the chiral condensate provides additional evidence for the nature of the phase transition. For example, if the transition were first-order, one would expect the histogram of the chiral condensate to develop a double-peak structure at the transition point. In~\autoref{fig:histogram_pbp_Nt12}, we show normalized histograms of the chiral condensate for ensembles located closest to the transition mass points determined from the peaks of the disconnected chiral susceptibility for $N_t=8$ and $N_t=12$. These correspond to $(m_q+\mres)^{\overline{\mathrm{MS}}}(\mu=2\,\mathrm{GeV})\simeq 42$ and $3.6$ MeV, respectively.

No pronounced evidence of a double-peak structure is observed as the volume increases. Instead, the histograms of the chiral condensate behave like a Gaussian distribution across different volumes at the simulated mass points, providing further evidence for a crossover transition. 

\begin{figure}[!htbp]
 	\centering
 		\includegraphics[scale=0.65]{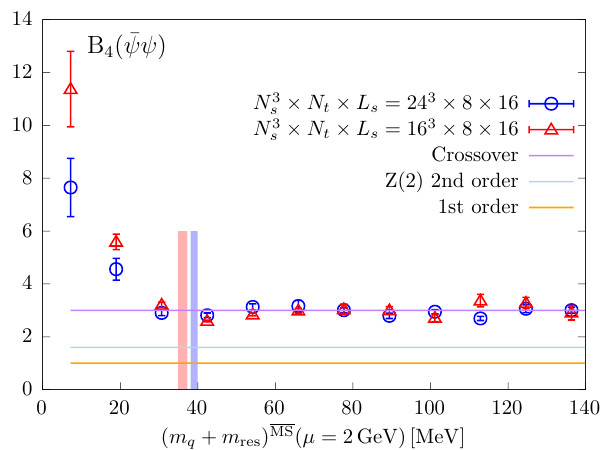}
 	\caption{Binder cumulant of the chiral condensate as a function of the renormalized quark mass evaluated on $N_s^3 \times 8 \times 16$ lattices with $N_s=16$ and 24. The horizontal lines represent the universal values for different types of phase transitions. The vertical bands represent the transition regions determined from the 1$\sigma$ error bounds (mean $\pm$ error) of the pseudocritical masses for the disconnected and total chiral susceptibilities, using only our largest volume ($24^3\times 8\times 16$).}
 	\label{fig:binder_cumulant_Nt8}
 \end{figure}

We also measured the Binder cumulant to determine the order of the phase transition at the transition mass point. In~\autoref{fig:binder_cumulant_Nt8}, we present the Binder cumulant
	 $B_4(\bar\psi \psi)$ evaluated on $16^3\times 8 \times 16$  and $24^3\times 8 \times 16$ lattices at $\beta=4.0$ as a function of the renormalized quark mass. The vertical bands indicate the transition regions at this fixed temperature, representing the 1$\sigma$ error bounds of the pseudocritical masses determined from both the disconnected and the total chiral susceptibilities for our largest volume ($24^3\times  8 \times 16$). 
The value of $B_4$ in these transition regions is consistent with 3, 
indicating a crossover transition. 
This conclusion aligns with the finding obtained from the volume independence of the chiral condensate and chiral susceptibilities,
 as illustrated in Figs.~\ref{fig:Nt8_pbp}, \ref{fig:Nt8_sub_pbp}, and \ref{fig:chi_disc_Nt8}.

\begin{figure}[!htbp]
	\centering
	\includegraphics[scale=0.65]{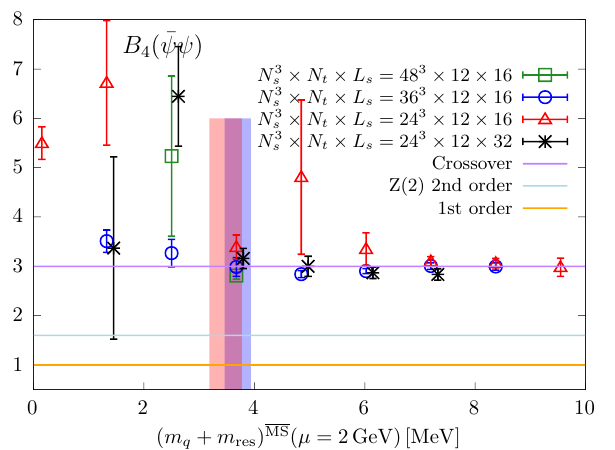}
	\caption{The Binder cumulant as a function of the renormalized quark mass for $N_t=12$ lattices with $L_s=16$ and $L_s=32$. The vertical bands indicate the transition mass regions, representing the 1$\sigma$ error bounds (mean$\pm$error) of the pseudocritical masses determined from the disconnected and total chiral susceptibilities for our largest volume ($48^3\times12\times16$).
	 }
	\label{fig:B4_Nt12}
\end{figure}

In \autoref{fig:B4_Nt12}, we present a plot similar to~\autoref{fig:binder_cumulant_Nt8}, but for $N_t=12$ lattices with $L_s=16$ and $L_s=32$. The vertical bands 
indicate the transition mass regions, representing the $1\sigma$ error bounds of the pseudocritical masses determined from both the disconnected and total chiral susceptibilities for our largest volume ($48^3\times 12\times 16$). Within these regions, the value of $B_4(\bar\psi \psi)$ is found to be close to 3 at $T = 121(2)$ MeV. This further supports the crossover scenario.

\section{Summary and outlook}
\label{label5}
We initiated a pioneering investigation of the nature of the three-flavor QCD phase transition toward the chiral limit
 using M\"{o}bius domain-wall fermions. We performed simulations at a fixed lattice spacing, $a=0.1361(20)$ fm, on lattices with temporal extents $N_t=6,8$, and 12, corresponding to temperatures of 242(4), 181(3), and 121(2) MeV, respectively. At each temperature, we explored a wide range of quark masses and several spatial volumes.  By evaluating the plaquette, plaquette susceptibility, chiral condensate, chiral susceptibilities, and Binder cumulant at different quark masses and volumes, we have estimated the pseudocritical quark masses at the two higher
 temperatures. On our largest simulated volumes, the pseudocritical mass at
 $T=242(4)$~MeV, determined from the plaquette susceptibility, is
  \[(m_q+m_{\rm res})^{\overline{\mathrm{MS}}}(\mu=2\,\mathrm{GeV}) \simeq 184(10)\, \mathrm{MeV},\]
  corresponding to $m_\pi\simeq0.96$~GeV,
 while at $T=181(3)$~MeV the peaks of the
 disconnected and total chiral susceptibilities give
 \[(m_q+m_{\rm res})^{\overline{\mathrm{MS}}}(\mu=2\,\mathrm{GeV}) \simeq 39.1(9)\,\mathrm{MeV}
 \quad\text{and}\quad 36(1)\,\mathrm{MeV},\]
i.e.\ $m_\pi\simeq445$ and $427$~MeV, respectively. These pseudocritical points lie unequivocally within the crossover region.
  
For $T=121(2)$ MeV, the pseudocritical quark masses determined from the peaks of the disconnected and total chiral susceptibilities on the largest volume are
\[ (m_q+m_{\mathrm{res}})^{\overline{\mathrm{MS}}} (\mu=2~\mathrm{GeV}) \simeq 3.7(2)~\mathrm{MeV} \quad\text{and}\quad 3.5(3)~\mathrm{MeV}, \]
i.e.\ $m_\pi\simeq137$ and $133$~MeV,
respectively. In contrast to the behavior at $T=181(3)$ MeV, 
where no significant volume dependence is observed, the peak height of the total chiral susceptibility at $T=121(2)$ MeV increases with increasing volume. However, this growth is substantially weaker than the power-law behavior anticipated for either a first-order or a $Z(2)$ second-order phase transition. The chiral condensate histograms and Binder cumulants also do not show definitive signatures of a true phase transition. Instead, the volume dependence of the susceptibility peak is well described by an exponential ansatz with saturation at large volumes, which is characteristic of a crossover. Nevertheless, since the peak height has not fully saturated across our simulated volumes, the possibility of a weak first-order phase transition cannot be completely ruled out. Our results therefore suggest that, if a first-order region exists in the lower-left corner of the Columbia plot, the corresponding $Z(2)$ critical boundary mass is likely below $3.7(2)$ MeV, close to the physical light-quark
mass~\cite{FlavourLatticeAveragingGroupFLAG:2021npn}, i.e.\
$m_\pi\lesssim137$~MeV. This finding is consistent with results obtained using Wilson and staggered-type fermions.

As a cross-check of the $N_t=12$, $L_s=16$ ensembles with negative input quark masses, we performed additional simulations on $24^3\times 12\times 32$ lattices. Comparing the $L_s=16$ and $L_s=32$ results, we find that the residual chiral symmetry breaking contribution to the chiral condensate is reduced at larger $L_s$, as expected, while the chiral susceptibilities remain consistent when compared at approximately the same total quark mass, $m_q + \mres$. This indicates that the chiral susceptibilities are primarily controlled by the total quark mass, rather than by the individual values of $m_q$ and $m_{\rm res}$.

From fits to the zero- and finite-temperature chiral-condensate data, we extract the coefficients $C^D$ and $x$ that enter the additive UV-divergent term $C^D(m_q+xm_{\rm res})/a^2$. We then explicitly subtract this contribution from the domain-wall-fermion chiral condensate. To our knowledge, this is the first explicit determination of both $C^D$ and $x$, and the first explicit subtraction of the additive UV-divergent contribution for domain-wall fermions. After this subtraction, the renormalized subtracted chiral condensates from the $L_s = 16$ and $L_s = 32$ ensembles agree when compared at the same total quark mass. Together with the consistency of the chiral susceptibilities, these observations support the robustness of the large negative input quark masses used for the $N_t = 12, L_s = 16$ ensembles.

In the case of a crossover transition, the peak height of the chiral susceptibilities should be volume independent for large volumes. For $N_t=12$ lattices, however, we have observed significant volume dependence in the chiral susceptibilities, though the effect is not as strong as one would expect for a true phase transition. To further clarify this ambiguity at $N_t=12$, future studies will require simulations on even larger spatial volumes (e.g., $N_s \ge 64$) to determine whether the peak height ultimately saturates in the thermodynamic limit, confirming a crossover, or continues to diverge, signaling a potentially weak true phase transition.
 Furthermore, to map out the broader phase structure and
directly search for the first-order region, 
if it exists, exploring even lower temperatures corresponding to a lighter quark mass transition point remains a critical direction. In particular, extending these M\"obius domain-wall fermion simulations to $N_t=14$ lattices, which would correspond to a temperature of 104(2) MeV, would be highly valuable. Such studies would provide an essential cross-check of a recent HISQ investigation that suggested  
 a continuous chiral phase transition with a transition temperature of $98^{+3}_{-6}$ MeV at a finite lattice spacing corresponding to $N_t=8$~\cite{Dini:2021hug}, and would further sharpen the constraints on the lower-left corner of the Columbia plot.

\section*{Acknowledgements}

We would like to thank Hidenori Fukaya and our other collaborators in the JLQCD collaboration for helpful discussions. 
This research benefitted from the computational resources made available by several institutions and projects, including Supercomputer Fugaku provided by the RIKEN Center for Computational Science through the 
HPCI projects hp220233 and hp210032, as well as Usability Research ra000001. Additionally, the work utilized the Wisteria/BDEC-01 Supercomputer System at Tokyo University/JCAHPC through the HPCI project hp220108 and the Ito supercomputer at Kyushu University through HPCI projects hp190124 and hp200050. The Hokusai BigWaterfall at RIKEN was also a valuable resource in this study. This work is also supported in part by JSPS KAKENHI Grant Numbers 20H01907, 
23K20846, and 25K01007. 
Y.~Zhang and J.~Goswami acknowledge support by the Deutsche Forschungsgemeinschaft (DFG, German Research Foundation) through the CRC-TR~211 ``Strong-interaction matter under extreme conditions''--project number 315477589--TRR~211.
We acknowledge the Grid Lattice QCD framework, \url{https://github.com/paboyle/Grid}, and its extension optimized for A64FX processors~\cite{Meyer:2019gbz}, which played a crucial role in generating the QCD configurations for this research. We thank N. Meyer and T. Wettig for their valuable discussions regarding the use of Grid for A64FX. For measurements, we used the Bridge++~\cite{Ueda:2014rya} and Hadrons codes~\cite{https://doi.org/10.5281/zenodo.4293902}, which significantly contributed to the success of this work.

\section*{Data Availability}
The numerical data that support the findings of this article are openly available in Zenodo~\cite{zhang_2026_21034066}.

\appendix
\section{The M\"{o}bius domain wall fermion}

\label{sec:appendix_action}
In this work, we employ the M\"{o}bius domain-wall fermion action, 
expressed in terms of 
five-dimensional M\"{o}bius domain-wall fermion fields $\Psi$ and  $\bar\Psi$ as
~\cite{Brower:2012vk}: 
\begin{align}
	\label{eq:MDWF}
	& S_{\text{MDWF}}(\bar\Psi, \Psi, U) = \bar\Psi D_{\text{MDWF}}(m_q) \Psi 
	\nonumber \\ &  \qquad\qquad  \qquad\quad = \sum_{s=0}^{L_s-1} \bar\Psi_s D^{(s)}_{+}\Psi_s + \sum_{s=1}^{L_s-1} \bar\Psi_s D^{(s)}_{-} P_{+} \Psi_{s-1}  + \sum_{s=0}^{L_s-2} \bar\Psi_s D^{(s)}_{-} P_{-}\Psi_{s+1}   \nonumber \\ &  \qquad\qquad  \qquad\quad - m_q\bar\Psi_0 D^{(0)}_{-} P_{+}\Psi_{L_s-1} - m_q \bar{\Psi}_{L_s-1} D^{(L_s-1)}_{-}P_{-}\Psi_0\,,
\end{align}
where $L_s$ is the size of the fifth dimension, and $m_q$ is the input bare quark mass. Here, $D^{(s)}_{+} = b_5(s) D^{\text{Wilson}}(M_5) + 1$, $D^{(s)}_{-} = c_5(s) D^{\text{Wilson}}(M_5) -1$, and $P_{\pm} = \frac{1}{2}(\mathbb{1} \pm \gamma_5)$. The four-dimensional Wilson-Dirac operator is defined as:
\begin{equation}
	\label{eq:wilson_dirac_operator}
	D_{x,y}^{\text{Wilson}}(U,M_5) = (4+M_5)\delta_{x,y} - \frac{1}{2} \left[ (1- \gamma_{\mu}) U_{\mu}(x) \delta_{x+\mu,y} + (1+\gamma_{\mu}) U^{\dagger}_{\mu}(y) \delta_{x, y+\mu} \right]\,,
\end{equation}
where $M_5$ is the domain-wall height.

Considering the polar decomposition to the sign function, it suffices to select constant coefficients $b_5(s) = b_5$ and $c_5(s)=c_5$. Using the real M\"{o}bius transformation of the Wilson-Dirac operator, the M\"{o}bius kernel can be expressed as:
\begin{equation}
\label{eq:mobius}
D^{\text{M\"obius}}(M_5) = \frac{(b_5 + c_5) D^{\text{Wilson}}(M_5)}{2 + (b_5 - c_5)D^{\text{Wilson}}(M_5)} 
\end{equation}
When $b_5 = a_5$ and $c_5=0$, the M\"{o}bius kernel reduces to the Shamir kernel, so
\begin{equation}
\label{eq:shamir}
D^{\text{Shamir}}(M_5) = \frac{a_5 D^{\text{Wilson}}(M_5)}{2 + a_5D^{\text{Wilson}}(M_5)} 
\end{equation}
If we consider the Shamir parameter $a_5$ fixed such that $a_5 = b_5 -c_5$, then the M\"{o}bius kernel is just a rescaled version of the Shamir kernel with a scaling factor $\alpha=(b_5 + c_5)/a_5$, 
\begin{equation}
\label{eq:mobius-shamir}
D^{\text{M\"obius}}(M_5) = \alpha D^{\text{Shamir}}(M_5)
\end{equation}
The M\"obius domain-wall fermion formulation has smaller residual chiral symmetry breaking effects than the Shamir domain-wall fermion. As demonstrated in~\cite{Brower:2012vk}, the scaling factor $\alpha$ exponentially enhances the reduction of the residual chiral symmetry breaking effect.

The M\"obius domain-wall fermion action with degenerate quark masses is invariant under a global $U(N_f)$ flavor symmetry. This symmetry allows for the definition of a five-dimensional conserved vector current: 
\begin{equation}
	\label{eq:5d_current_four_component}
	\begin{aligned}
		j_\mu^{a, \text{MDWF}}(x, s) = & \, b_5 \bar{\Psi}_s \lambda^a \mathcal{V}_\mu(x) \Psi_s 
		+ c_5 \bar{\Psi}_s \lambda^a \mathcal{V}_\mu(x) P_{-} \Psi_{s+1} 
		+ c_5 \bar{\Psi}_s \lambda^a  \mathcal{V}_\mu(x)P_{+} \Psi_{s-1} \\
		& - c_5 (1 + m_q) \bar{\Psi}_0 \lambda^a  \mathcal{V}_\mu(x) P_{+} \Psi_{L_s-1} \, \delta_{s,0} \\
		& - c_5 (1 + m_q) \bar{\Psi}_{L_s-1} \lambda^a\mathcal{V}_\mu(x) P_- \Psi_0 \, \delta_{s,L_s-1}\,,  
	\end{aligned}
\end{equation}
for $\mu = 1, \dots, 4$, and $0 \le s \le  L_s-1$. The fifth component of the current is:
\begin{equation}
	\label{eq:5d_current_fifth_component}
	j_5^{a, \text{MDWF}}(x, s) =
	\begin{cases}
		\, \bar{\Psi}_{x, s+1}  \lambda^a D_-^{(s+1)} P_+ \Psi_{x, s} - \bar{\Psi}_{x,s} \lambda^a D_-^{(s)} P_- \Psi_{x, s+1}, & 0 \leq s < L_s-1, \\
		\bar{\Psi}_{x, 0} \lambda^a D_-^{(0)} P_+ \Psi_{x, L_s-1} -\bar{\Psi}_{x, L_s-1} \lambda^a D_-^{(L_s-1)} P_- \Psi_{x, 0}, & s = L_s-1.
	\end{cases}
\end{equation}
Here, $	\mathcal{V}_\mu(x) = \frac{\gamma_\mu - 1}{2} U_\mu(x)  + \frac{\gamma_\mu + 1}{2} U_\mu^\dagger(x)$.
The five-dimensional theory is anomaly-free, so the divergence equation satisfies:
\begin{equation}
\label{eq:divergence_eq}
\sum_{\mu=1}^{4}\Delta_{\mu} 	j_{\mu}^{a, \text{MDWF}}(x, s) + \Delta_{5} 	j_5^{a, \text{MDWF}}(x, s) =0
\end{equation}
Written out explicitly, this becomes:
\begin{equation}
\sum_{\mu=1}^{4} \Delta_\mu j_\mu^{a, \text{MDWF}}(x, s) = 
\begin{cases} 
-j_5^a(x, 0) - m_q j_5^a(x, L_s-1), & s = 0, \\[8pt]
-\Delta_5 j_5^{a}(x, s), & 0 < s < L_s-1, \\[8pt]
j_5^a(x, L_s-2) + m_q j_5^a(x, L_s-1), & s = L_s-1.
\end{cases}
\end{equation}
Here,
\begin{equation}
\Delta_5 f(x, s)  = f(x, s) - f(x, s-1)
\end{equation}	
One then defines a conserved four-dimensional vector current by summing over the fifth dimension
\begin{equation}
J^{a, \text{MDWF}}_\mu(x) = \sum_{s=0}^{L_s-1} j_\mu^{a, \text{MDWF}}(x, s) 
\end{equation}
since no current escapes through the fifth dimension, 
\begin{equation}
\sum_{s=0}^{L_s-1} \Delta_{5} 	j_5^{a, \text{MDWF}}(x, s) =0
\end{equation}

The four-dimensional axial current is defined by splitting the left- and right-handed currents 
in the fifth dimension with opposite signs: 
\begin{equation}
	\label{eq:axial_current}
	\mathcal{A}^a_{\mu}(x) = \sum_{s=0}^{L_s-1} \operatorname{sign}(s - \frac{L_s-1}{2})j_\mu^{a, \text{MDWF}}(x, s) 
\end{equation}
The divergence of the four-dimensional axial current satisfies: 
\begin{equation}
	\label{eq:axial_current_div}
	\sum_{\mu=1}^{4}\Delta_{\mu} \mathcal{A}^a_{\mu}(x) = 2m_q J_5^{a, \text{MDWF}}(x) + 2  J^{a, \text{MDWF}}_{5q}(x)
\end{equation}
where  $J_{5}^{a,\mathrm{MDWF}}(x)$ and $J_{5q}^{a,\mathrm{MDWF}}(x)$ are the pseudoscalar densities defined in Eqs.~\eqref{eq:J_5} and \eqref{eq:J_5q}, respectively.

Under the axial current transformation, the change in the M\"obius domain-wall fermion action comes from two sides: one is the coupling between $s=L_s/2 -1$ and $s=L_s/2$ layers, 
which persists 
even when $m_q=0$, and the other one comes from the coupling between $s=0$ and $s=L_s-1$, 
which vanishes when $m_q=0$.

The corresponding axial Ward-Takahashi identity is~\cite{Furman:1994ky}: 
\begin{equation}
	\label{eq:WTI}
	\sum_{\mu=1}^{4}\Delta_{\mu}\langle \mathcal{A}^a_{\mu}(x) O(y) \rangle = 2m_q\langle J_5^a(x)O(y)\rangle + 2 \langle  J^a_{5q}(x)O(y)\rangle +  i\langle \delta^aO(y)\rangle 
\end{equation}
It is shown in~\cite{Furman:1994ky} that if the operator $O$ only involves the four-dimensional quark field as defined in Eqs.~\eqref{eq:q_x} and~\eqref{eq:qbarx}, then the correlation function $\langle  J^a_{5q}(x)O(y)\rangle$ will tend to zero as $L_s \to \infty$. This implies that as the scale of the fifth dimension tends to infinity, we can obtain the expected Ward identity.

When $O(y) = J^b_5(y) = \bar{\psi}_y \gamma_5 \lambda^b \psi_y$,  
the corresponding Ward identity is:
\begin{align}
	\begin{split}
		\label{eq:ward-identity}
		\sum_{\mu=1}^{4} \Delta_{\mu}  \langle \mathcal{A}_{\mu}^a(x) J_5^b(y) \rangle  & = 2m_q  \langle J_{5}^a(x) J_5^b(y) \rangle  + 2 \langle J_{5q}^a(x) J_5^b(y) \rangle - \delta_{x,y} \langle \bar{\psi}_y\{\lambda^a, \lambda^b\} \psi_y\rangle   \,,
	\end{split}
\end{align}
where $\lambda^a$ is a flavor symmetry generator, and $\{ \lambda^a, \lambda^b\} = 2\delta^{ab}\mathbb{1}$.

\begin{figure}[!htp]
	\centering
	\includegraphics[scale=0.8]{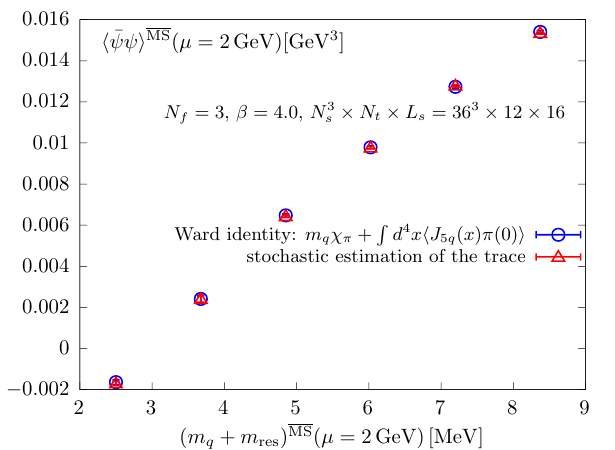}
\caption{Numerical verification of the integrated axial Ward--Takahashi 
			identity as shown in Eq.~\eqref{eq:ward} for $36^3\times 12\times 16$ lattices. The renormalized chiral condensate 
			obtained using the Ward identity $m_q\chi_\pi + \int d^4x\,\langle J_{5q}^a(x)\pi(0)\rangle$ (circles) and stochastic estimator (triangles) are shown as a function of the renormalized quark mass.}
	\label{fig:ward}
\end{figure}

Summing over all lattice points gives~\cite{PhysRevD.69.074502,PhysRevD.89.054514}:
\begin{align}
	\begin{split}
		\label{eq:integrated-ward-identity}
		0 & = \int d^4 x\,\sum_{\mu=1}^{4}	\Delta_{\mu}  \langle \mathcal{A}_{\mu}^a(x) J_5^b(y) \rangle = \int d^4 x\, \left(2m_q  \langle J_{5}^a(x) J_5^b(y) \rangle  + 2 \langle J_{5q}^a(x) J_5^b(y) \rangle - \delta_{x,y} \langle 
		\bar{\psi}_y\{\lambda^a, \lambda^b\} \psi_y\rangle \right)  \\
		& = 2m_q \chi_{\pi} + 2	\Delta _{\mathrm {mp}} - 2  \langle \bar \psi \psi \rangle\,,
	\end{split}
\end{align}
yielding
\begin{equation}
	\label{eq:ward}
	m_q\chi_{\pi}+ \Delta_{\mathrm {mp}} = \langle \bar \psi \psi \rangle  \,,
\end{equation}
which we will refer to as the integrated axial Ward--Takahashi identity. Here,
\begin{equation}
\label{eq:delta_mp}
\Delta _{\mathrm {mp}} = \int d^4 x\,\langle J^a_{5q}(x) J_5^b(y) \rangle \,.
\end{equation}
In Fig.~\ref{fig:ward}, we numerically verify the integrated axial Ward--Takahashi identity, Eq.~\eqref{eq:ward}, by comparing two independent determinations of the chiral condensate: one obtained from the Ward identity and the other from the stochastic estimator. The excellent agreement between the two determinations confirms the validity of the identity.
\FloatBarrier
\section{Details on simulation parameters}
\label{app:params}

In Tables~\ref{sup:table_nt8} and \ref{sup:table_nt12} we also quote the
pseudoscalar mass, obtained from $m_\pi^2 = 2B\,(m_q+m_{\rm res})$ with
$am_{\rm res}=0.00613$ for $L_s=16$ and $0.00324$ for $L_s=32$, and $2B$
determined in the same convention from the $\beta=4.0$ zero-temperature data of
Fig.~\ref{fig:mpi_mass}, constrained to pass through the origin. These values are
leading-order estimates meant as guidance; no value is quoted when the
renormalized quark mass is consistent with zero.

\begin{table}[htb]
	\scriptsize
	\centering
	\renewcommand{\arraystretch}{1.05}
	\begin{tabular}{c|c|c||c|c|c}
		\hline
		\hline
		\makebox[1.2cm][c]{$am_q$} & \makebox[1.4cm][c]{$\#$streams} & \makebox[1.4cm][c]{$\#$conf} &
		\makebox[1.2cm][c]{$am_q$} & \makebox[1.4cm][c]{$\#$streams} & \makebox[1.4cm][c]{$\#$conf} \\
		\hline
		\hline
		\multicolumn{6}{c}{$12^3 \times 6 \times 16$} \\
		\hline
		0.000 & 8 & 10336 & 0.210 & 8 & 10420 \\
		0.010 & 8 & 10300 & 0.220 & 8 & 10431 \\
		0.020 & 8 & 10434 & 0.230 & 8 & 10419 \\
		0.030 & 8 & 10468 & 0.240 & 8 & 10337 \\
		0.040 & 8 & 10399 & 0.250 & 8 & 10257 \\
		0.050 & 8 & 10398 & 0.260 & 8 & 10397 \\
		0.060 & 8 & 10426 & 0.270 & 8 & 10243 \\
		0.070 & 8 & 10277 & 0.280 & 8 & 10129 \\
		0.080 & 8 & 10326 & 0.290 & 8 & 10377 \\
		0.090 & 8 & 10440 & 0.300 & 8 & 10350 \\
		0.100 & 8 & 10398 & 0.310 & 8 & 10115 \\
		0.110 & 8 & 10402 & 0.320 & 8 & 10425 \\
		0.120 & 8 & 10454 & 0.330 & 8 & 10466 \\
		0.130 & 8 & 10374 & 0.340 & 8 & 10401 \\
		0.140 & 8 & 10428 & 0.350 & 8 & 10492 \\
		0.150 & 8 & 10440 & 0.360 & 8 & 10280 \\
		0.160 & 8 & 10228 & 0.370 & 8 & 10438 \\
		0.170 & 8 & 10421 & 0.380 & 8 & 10422 \\
		0.180 & 7 & 9097  & 0.390 & 8 & 10479 \\
		0.190 & 6 & 7657  & 0.400 & 8 & 10441 \\
		0.200 & 8 & 10383 &       &   &       \\
		\hline
		\hline
		\multicolumn{6}{c}{$16^3 \times 6 \times 16$} \\
		\hline
		0.050 & 10 & 9907 & 0.160 & 10 & 8910 \\
		0.060 & 10 & 9910 & 0.170 & 10 & 8027 \\
		0.070 & 10 & 9410 & 0.180 &  8 & 7119 \\
		0.080 & 10 & 8910 & 0.190 &  8 & 7120 \\
		0.090 & 10 & 8910 & 0.200 & 10 & 8910 \\
		0.100 & 10 & 8910 & 0.210 & 10 & 8906 \\
		0.110 & 10 & 8910 & 0.220 & 10 & 8910 \\
		0.120 & 10 & 8910 & 0.230 & 10 & 9010 \\
		0.130 & 10 & 8910 & 0.240 & 10 & 9110 \\
		0.140 & 10 & 8910 & 0.250 & 10 & 9810 \\
		0.150 & 10 & 8910 &       &    &      \\
		\hline
		\hline
	\end{tabular}
	\caption{Simulation parameters and statistics for the $N_t=6$
		finite-temperature ensembles at $\beta=4.0$, corresponding to
		$T=242(4)$~MeV. Each quark mass was simulated with several independent
		streams; the first 1000 trajectories of each stream were discarded for thermalization, and configurations were sampled every 10 trajectories.}
	\label{sup:table_nt6}
\end{table}

\begin{table}[htb]
	\scriptsize
	\centering
	\renewcommand{\arraystretch}{1.08}
	\begin{tabular}{c|c|c|c|c|c}
		\hline
		\hline
		\makebox[1.5cm][c]{$\beta$} &
		\makebox[2.0cm][c]{$T$ (MeV)} &
		\makebox[2.8cm][c]{$N_s^3 \times N_t \times L_s$} &
		\makebox[1.5cm][c]{$am_q$} &
		\makebox[1.8cm][c]{$m_\pi$ (MeV)} &
		\makebox[1.5cm][c]{$\#$conf} \\
		\hline
		\hline
		\multirow{11}{*}{4} & \multirow{11}{*}{181(3)} & \multirow{11}{*}{$16^3 \times 8 \times 16$} & 0 & 191 & 95 \\
		& & & 0.02 & 394 & 106 \\
		& & & 0.04 & 523 & 98 \\
		& & & 0.06 & 627 & 108 \\
		& & & 0.08 & 715 & 113 \\
		& & & 0.10 & 794 & 116 \\
		& & & 0.12 & 865 & 115 \\
		& & & 0.14 & 932 & 119 \\
		& & & 0.16 & 993 & 124 \\
		& & & 0.18 & 1051 & 125 \\
		& & & 0.20 & 1106 & 123 \\
		\hline
		\multirow{15}{*}{4} & \multirow{15}{*}{181(3)} & \multirow{15}{*}{$24^3 \times 8 \times 16$} & 0 & 191 & 51 \\
		& & & 0.01 & 310 & 56 \\
		& & & 0.02 & 394 & 42 \\
		& & & 0.03 & 463 & 41 \\
		& & & 0.04 & 523 & 60 \\
		& & & 0.05 & 577 & 60 \\
		& & & 0.06 & 627 & 60 \\
		& & & 0.07 & 672 & 60 \\
		& & & 0.08 & 715 & 60 \\
		& & & 0.09 & 756 & 60 \\
		& & & 0.10 & 794 & 57 \\
		& & & 0.11 & 830 & 60 \\
		& & & 0.12 & 865 & 60 \\
		& & & 0.13 & 899 & 59 \\
		& & & 0.14 & 932 & 60 \\
		\hline
		\hline
	\end{tabular}
		\caption{Simulation parameters and statistics for the finite-temperature ensembles at $N_t=8$. The lattice spacing is fixed at $a = 0.1361(20)$ fm, with the temperature determined via $T = 1/(N_t a)$.
			The first 1100 trajectories were discarded for thermalization. Configurations were sampled every 100 trajectories across 10 streams.}
	\label{sup:table_nt8}
\end{table}

\begin{table}[htb]
	\scriptsize
	\centering
	\renewcommand{\arraystretch}{1.05}
	\begin{tabular}{c|c|c|c|c|c}
		\hline
		\hline
		\makebox[1.5cm][c]{$\beta$} &
		\makebox[2.0cm][c]{$T$ (MeV)} &
		\makebox[2.8cm][c]{$N_s^3 \times N_t \times L_s$} &
		\makebox[1.5cm][c]{$am_q$} &
		\makebox[1.8cm][c]{$m_\pi$ (MeV)} &
		\makebox[1.5cm][c]{$\#$conf} \\
		\hline
		\hline
		\multirow{26}{*}{4.0} & \multirow{26}{*}{121(2)} & \multirow{26}{*}{$24^3 \times 12 \times 16$} &  0.1 & 794 & 2452 \\
		& & & 0.090 & 756 & 2374 \\
		& & & 0.080 & 715 & 2361 \\
		& & & 0.070 & 672 & 2332 \\
		& & & 0.060 & 627 & 2063 \\
		& & & 0.050 & 577 & 2033 \\
		& & & 0.040 & 523 & 2034 \\
		& & & 0.030 & 463 & 1758 \\
		& & & 0.020 & 394 & 1604 \\
		& & & 0.010 & 310 & 1366 \\
		& & & 0.009 & 300 & 1357 \\
		& & & 0.008 & 290 & 1329 \\
		& & & 0.007 & 279 & 1322 \\
		& & & 0.006 & 268 & 1330 \\
		& & & 0.005 & 257 & 1076 \\
		& & & 0.004 & 245 & 1116 \\
		& & & 0.003 & 233 & 984 \\
		& & & 0.002 & 220 & 1023 \\
		& & & 0.001 & 206 & 1628 \\
		& & & 0 & 191 & 1445 \\
		& & & -0.001 & 175 & 631 \\
		& & & -0.002 & 157 & 3183 \\
		& & & -0.003 & 136 & 3032 \\
		& & & -0.004 & 112 & 2772 \\
		& & & -0.005 & 82 & 2238 \\
		& & & -0.006 & --- & 2057 \\
		\hline
		\multirow{7}{*}{4.0} & \multirow{7}{*}{121(2)} & \multirow{7}{*}{$36^3 \times 12 \times 16$} &  0.001 & 206 & 3734 \\
		& & & 0 & 191 & 4034 \\
		& & & -0.001 & 175 & 4307 \\
		& & & -0.002 & 157 & 3859 \\
		& & & -0.003 & 136 & 2173 \\
		& & & -0.004 & 112 & 2229 \\
		& & & -0.005 & 82 & 722 \\
		\hline
		\multirow{2}{*}{4.0} & \multirow{2}{*}{121(2)} & \multirow{2}{*}{$48^3 \times 12 \times 16$} &  -0.003 & 136 & 1915 \\
		& & & -0.004 & 112 & 1986 \\
		\hline
		\multirow{5}{*}{4.0} & \multirow{5}{*}{121(2)} & \multirow{5}{*}{$24^3 \times 12 \times 32$} &  0.003 & 192 & 1786 \\
		& & & 0.002 & 176 & 1656 \\
		& & & 0.001 & 159 & 1870 \\
		& & & 0 & 139 & 1641 \\
		& & & -0.001 & 115 & 1614 \\
		\hline
		\hline
	\end{tabular}
	\caption{Simulation parameters and statistics for $N_t=12$ finite-temperature ensembles. The first 500-800 trajectories were discarded for thermalization. Configurations are stored every 10 trajectories.}
	\label{sup:table_nt12}
\end{table}

\begin{table}[htb]
	\small
	\centering
	\renewcommand{\arraystretch}{1.05}
	\begin{tabular}{c|c|c|c|c|c}
		\hline
		\hline
		\makebox[1.5cm][c]{$\beta$} &
		\makebox[2.0cm][c]{$a^{-1}$ (GeV)} &
		\makebox[2.8cm][c]{$N_s^3 \times N_t \times L_s$} &
		\makebox[1.5cm][c]{$am_q$} &
		\makebox[1.8cm][c]{$m_\pi$ (MeV)} &
		\makebox[1.5cm][c]{$\#$conf} \\
		\hline
		\hline
			\multirow{3}{*}{4.0} & \multirow{3}{*}{1.449} & \multirow{3}{*}{$12^3 \times 24 \times 16$} & 0.1 & 822.7(22) & 69 \\
& & & 0.2 & 1189.0(10) & 69 \\
& & & 0.3 & 1491.1(9) & 70 \\
		\hline
		\multirow{6}{*}{4.0} & \multirow{6}{*}{1.449} & \multirow{6}{*}{$24^3 \times 48 \times 16$} & 0.020 & 391.1(5) & 78 \\
		& & & 0.025 & 428.5(9) & 95 \\
		& & & 0.030 & 461.7(8) & 100 \\
		& & & 0.035 & 494.5(8) & 152 \\
		& & & 0.040 & 525.4(6) & 109 \\
		& & & 0.045 & 554.8(9) & 109 \\
		\hline
		\multirow{6}{*}{4.1} & \multirow{6}{*}{2.033} & \multirow{6}{*}{$24^3 \times 48 \times 16$} & 0.015 & 398.9(32) & 362 \\
		& & & 0.020 & 456.3(20) & 427 \\
		& & & 0.025 & 504.4(23) & 457 \\
		& & & 0.030 & 554.2(14) & 467 \\
		& & & 0.035 & 598.1(16) & 511 \\
		& & & 0.040 & 641.5(17) & 539 \\
		\hline
		\multirow{6}{*}{4.17} & \multirow{6}{*}{2.453} & \multirow{6}{*}{$32^3 \times 64 \times 16$} & 0.012 & 386.5(22) & 166 \\
		& & & 0.015 & 434.4(20) & 181 \\
		& & & 0.018 & 482.3(17) & 190 \\
		& & & 0.021 & 520.5(16) & 220 \\
		& & & 0.024 & 555.6(14) & 247 \\
		& & & 0.026 & 580.1(17) & 253 \\
		\hline
	\end{tabular}
	\caption{Simulation parameters and statistics for the zero-temperature ensembles. The first 300-450 trajectories were discarded for thermalization. Configurations for the $12^3 \times 24 \times 16$ ensemble were sampled every 100 trajectories, while for all other ensembles, configurations were sampled every 10 trajectories. In contrast to Tables~\ref{sup:table_nt8} and \ref{sup:table_nt12}, the
		pseudoscalar masses quoted here are measured directly on these ensembles; the quoted errors are statistical.}
	\label{sup:table_zeroT}
\end{table}

\clearpage
\section{Reweighting of the rational functions}
\label{app:reweight}
\subsection{Rational approximation}
For simplicity, we  illustrate the reweighting of the rational functions for the case of a single flavor of domain-wall fermions. The pseudofermion action for one flavor, 
approximated by a rational function, reads
\begin{align}
	S_{pf}  & =  \phi^\dagger \mathcal{D}(1)^{1/4} \mathcal{D}(m_q)^{-1/2} \mathcal{D}(1)^{1/4} \phi \label{eq:C1}\\
&	\approx \phi^\dagger \left( a_0 +  \sum_i \frac{a_i}{\mathcal{D}(1) + b_i} \right) \left( \tilde{a}_0 + \sum_i \frac{\tilde{a}_i}{\mathcal{D}(m_q) + \tilde{b}_i} \right) \left(  a_0 + \sum_i \frac{a_i}{\mathcal{D}(1) + b_i} \right) \phi\,, \label{eq:C2}
\end{align}
where $\mathcal{D}(m_q) = D^{\dagger}_{\text{MDWF}}(m_q)D_{\text{MDWF}}(m_q)$ 
is the hermitian two-flavor M\"{o}bius domain-wall Dirac operator with 
fermion mass $m_q$. The coefficients $\left(a_0, a_i, b_i\right)$ 
and $\left(\tilde{a}_0, \tilde{a}_i, \tilde{b}_i\right)$ are the rational 
approximation coefficients for $\mathcal{D}(1)^{1/4}$ and 
$\mathcal{D}(m_q)^{-1/2}$, respectively. Integrating out the 
pseudofermion fields yields the fermion determinant:
\begin{align}
	\label{integration}
\int D\phi^{\dagger} D\phi \, e^{-\phi^{\dagger} \mathcal{D}(1)^{1/4} \mathcal{D}(m_q)^{-1/2} \mathcal{D}(1)^{1/4}  \phi } =\det  \left( \mathcal{D}(1)^{-1/4} \mathcal{D}(m_q)^{1/2}  \mathcal{D}(1)^{-1/4} \right)
= {\det \left(\frac{\mathcal{D}(m_q)}{\mathcal{D}(1)}\right)}^{1/2} .
\end{align}
The pseudofermion field $\phi$ 
is generated from a Gaussian stochastic noise field  $\xi$ satisfying $\langle\xi_i^{\dagger}\xi_j\rangle=\delta_{ij}$ via  
\begin{equation}
	\phi = \mathcal{D}(1)^{-1/4}\mathcal{D}(m_q)^{1/4}\xi,
	\label{pseudofermion_field}
\end{equation}
which, in the rational approximation, reads
\begin{equation}
	\phi \approx \left(\alpha_0 + \sum_i \frac{\alpha_i}
	{\mathcal{D}(1)+\beta_i}\right)
	\left(\tilde{\alpha}_0 + \sum_i \frac{\tilde{\alpha}_i}
	{\mathcal{D}(m_q)+\tilde{\beta}_i}\right)\xi,
		\label{eq:C5}
\end{equation}
where $(\alpha_0, \alpha_i, \beta_i)$ and 
$(\tilde{\alpha}_0, \tilde{\alpha}_i, \tilde{\beta}_i)$ are the rational 
approximation coefficients for $\mathcal{D}(1)^{-1/4}$ and 
$\mathcal{D}(m_q)^{1/4}$, respectively.
Changing the integration variable from $\phi$ to $\xi$ via Eq.~\eqref{pseudofermion_field}
 in Eq.~\eqref{integration} then yields
\begin{equation}
	\int D\xi^\dagger D\xi \left(\det \frac{\mathcal{D}(m_q)}{\mathcal{D}(1)}\right)^{1/2} e^{-\xi^\dagger\xi}\,.
\end{equation}
For each operator, $\mathcal{D}(m_q)$ and $\mathcal{D}(1)$, we need two rational approximations. The rational approximations should satisfy the following conditions:
\begin{itemize}
	\item The rational approximation in Eq.~\eqref{eq:C2} should 
	accurately reproduce $\mathcal{D}(1)^{1/4}\mathcal{D}(m_q)^{-1/2}
	\mathcal{D}(1)^{1/4}$ in calculating the final Hamiltonian.
	\item For $\xi^\dagger\xi$ to give the initial pseudofermion 
	action, the approximation should satisfy the following condition, 
	obtained by substituting Eq.~\eqref{eq:C5} into Eq.~\eqref{eq:C1}:
\begin{align}
	1 = &\left(\tilde{\alpha}_0 + \sum_i \frac{\tilde{\alpha}_i}
	{\mathcal{D}(m_q) + \tilde{\beta}_i}\right)
	\left(\alpha_0 + \sum_i \frac{\alpha_i}
	{\mathcal{D}(1) + \beta_i}\right) \nonumber \\
	&\times \left(a_0 + \sum_i \frac{a_i}
	{\mathcal{D}(1) + b_i}\right)
	\left(\tilde{a}_0 + \sum_i \frac{\tilde{a}_i}
	{\mathcal{D}(m_q) + \tilde{b}_i}\right)
	\left(a_0 + \sum_i \frac{a_i}
	{\mathcal{D}(1) + b_i}\right) \nonumber \\
	&\times \left(\alpha_0 + \sum_i \frac{\alpha_i}
	{\mathcal{D}(1) + \beta_i}\right)
	\left(\tilde{\alpha}_0 + \sum_i \frac{\tilde{\alpha}_i}
	{\mathcal{D}(m_q) + \tilde{\beta}_i}\right).
	\label{eq:C7}
\end{align}
	\item During the molecular dynamics, the accuracy of the rational approximation is not very important as long as the reversibility is kept.
\end{itemize}

If the condition for the pseudofermion is violated, the reversibility is lost. It is expensive to monitor directly the reversibility during production runs, but we can monitor its violation by checking if $\langle\exp(-dH)\rangle=1$ is satisfied. If the approximation for the final Hamiltonian is poor, the distribution of the configuration is skewed. It should be accurate enough to resolve $dH$ in the Metropolis test.

The older version of Grid appears to use Eq.~(\ref{eq:C2}) instead of $\xi^\dagger \xi$ to calculate the initial action so that the reversibility is fine. But the poor accuracy of the approximation implies that the connection to the Gaussian distribution is poor, and the distribution of the pseudofermion becomes inaccurate.

There are three sources that cause poor accuracy in the rational approximation:
\begin{itemize}
	\item The eigenvalues spread beyond the support of the approximation. In the case of Grid, it sometimes checks the bound of the eigenvalues and RHMC stops if it goes outside of the support.
	\item On the support, the rational function approximates the target function within a certain finite precision. We need a sufficient number of terms to realize the precision we need.
	\item The precision of the solvers is also relevant.
\end{itemize}
Assuming the range of the eigenvalues is valid, the precision of the rational function or the solver governs the accuracy of the rational approximation. If the number of terms is not large enough but the solver is precise so that $dH$ makes sense, we can conclude that the RHMC generates configurations with a slightly different weight.

The effect of the solver precision is more involved in practice. Even if the precision of one term $(\mathcal{D}(1) + b_i)^{-1}$, for example, is poor, if the coefficient of this term $a_i$ is small its effect on the result is limited. For a given target tolerance, the solutions of the multi-shift solver tend to yield more accurate solutions for larger shift $b_i$. The coefficient $a_i$ in the rational function becomes larger as the shift $b_i$ becomes larger. As a result, one can expect better accuracy than the target tolerance passed to the multi-shift solver. 

Let us denote the rational function that approximates $x^p$ by $f_p(x)$ and a less accurate approximation by $f_p^{\mathrm{poor}}(x)$. We summarize the situation below.

What we want:
\begin{itemize}
	\item pseudofermion: $\phi = f_{-1/4}(\mathcal{D}(1)) f_{1/4}(\mathcal{D}(m_q)) \xi$
	\item initial action: $\xi^\dagger \xi$
	\item action for MD: $\phi^\dagger f_{1/4}(\mathcal{D}(1)) f_{-1/2}(\mathcal{D}(m_q)) f_{1/4}(\mathcal{D}(1)) \phi$
	\item final action: $\phi^\dagger f_{1/4}(\mathcal{D}(1)) f_{-1/2}(\mathcal{D}(m_q)) f_{1/4}(\mathcal{D}(1)) \phi$
\end{itemize}

What we actually did:
\begin{itemize}
	\item pseudofermion: $\tilde{\phi} = f^{\rm poor}_{-1/4}(\mathcal{D}(1)) f^{\rm poor}_{1/4}(\mathcal{D}(m_q)) \xi$
	\item initial action: $\xi^\dagger \xi$ or $\tilde{\phi}^\dagger f^{\rm poor}_{1/4}(\mathcal{D}(1)) f^{\rm poor}_{-1/2}(\mathcal{D}(m_q)) f^{\rm poor}_{1/4}(\mathcal{D}(1)) \tilde{\phi}$
	\item action for MD: $S^{\rm poor} = \tilde{\phi}^\dagger f^{\rm poor}_{1/4}(\mathcal{D}(1)) f^{\rm poor}_{-1/2}(\mathcal{D}(m_q)) f^{\rm poor}_{1/4}(\mathcal{D}(1)) \tilde{\phi}$
	\item final action: $\tilde{\phi}^\dagger f^{\rm poor}_{1/4}(\mathcal{D}(1)) f^{\rm poor}_{-1/2}(\mathcal{D}(m_q)) f^{\rm poor}_{1/4}(\mathcal{D}(1)) \tilde{\phi}$
\end{itemize}
\subsection{Reweighting to correct the rational function}
For $36^3\times 12\times 16$ runs, it turned out that the number of terms for the rational function is not large enough. The following numbers come from a run for $36^3 \times 12 \times 16$ configurations with 12 poles for the rational functions:
\begin{itemize}
	\item the errors are $1.387188049647018\times10^{-6}$ ($x^{1/2}$ and $x^{-1/2}$), and $9.662807343093718\times10^{-7}$ ($x^{1/4}$ and $x^{-1/4}$)
	\item the size of the pseudofermion action: $5.37\times10^{7}$
	\item the target solver tolerance: $1.0\times10^{-7}$
\end{itemize}

The observed $\Delta H$ distribution is reasonable and satisfies
$\langle e^{-\Delta H}\rangle \simeq 1$, indicating that the HMC evolution is valid for the approximate rational action. However, the finite error of the rational approximation slightly modifies the target distribution. We therefore apply rational-function reweighting to correct this difference.

An older version of the Grid code uses $S^{\rm poor}$ to estimate the initial action. We denote the relation between the correct pseudofermion $\phi$ and the one we actually used $\tilde{\phi}$ as
\begin{align}
	\tilde{\phi} = C \phi.
\end{align}
Then, both the initial and final action as well as the MD process use
\begin{align}
	\tilde{S}_C = \phi^\dagger C^\dagger f^{\rm poor}_{1/4}(\mathcal{D}(1)) f^{\rm poor}_{-1/2}(\mathcal{D}(m_q)) f^{\rm poor}_{1/4}(\mathcal{D}(1)) C \phi
\end{align}
and the fermion determinant
\begin{align}
	\det \left[ C^\dagger f^{\rm poor}_{1/4}(\mathcal{D}(1)) f^{\rm poor}_{-1/2}(\mathcal{D}(m_q)) f^{\rm poor}_{1/4}(\mathcal{D}(1)) C \right]^{-1}.
\end{align}
As
\begin{align}
	C = f^{\rm poor}_{-1/4}(\mathcal{D}(1)) f^{\rm poor}_{1/4}(\mathcal{D}(m_q)) f_{-1/4}(\mathcal{D}(m_q)) f_{1/4}(\mathcal{D}(1))\,,
\end{align}
the reweighting factor is
\begin{align}
	w &= \det \left[ \frac{C^\dagger 
		f^{\text{poor}}_{1/4}(\mathcal{D}(1))
		f^{\text{poor}}_{-1/2}(\mathcal{D}(m_q))
		f^{\text{poor}}_{1/4}(\mathcal{D}(1))C}
	{f_{1/4}(\mathcal{D}(1))
		f_{-1/2}(\mathcal{D}(m_q))
		f_{1/4}(\mathcal{D}(1))} \right] \nonumber \\
	&= \det \Big[ 
	f^{\text{poor}}_{-1/4}(\mathcal{D}(1))
	f^{\text{poor}}_{1/4}(\mathcal{D}(1))
	f^{\text{poor}}_{1/4}(\mathcal{D}(m_q))
	f^{\text{poor}}_{-1/2}(\mathcal{D}(m_q))
	f^{\text{poor}}_{1/4}(\mathcal{D}(m_q))
	f^{\text{poor}}_{1/4}(\mathcal{D}(1))
	f^{\text{poor}}_{-1/4}(\mathcal{D}(1))
	\Big] \nonumber \\
	&= \det \Omega .
	\label{eq:C12}
\end{align}
In general,
\begin{equation}
	\frac{1}{\det \Omega} = 
	\frac{\int d\xi \exp(-\xi^\dagger \Omega \xi)}
	{\int d\xi \exp(-\xi^\dagger \xi)}
	= \frac{\int d\xi \exp(-\xi^\dagger(\Omega - 1)\xi)
		\exp(-\xi^\dagger \xi)}
	{\int d\xi \exp(-\xi^\dagger \xi)},
	\label{eq:det_omega}
\end{equation}
therefore, using a Gaussian noise vector $\xi$, we can estimate 
the above as a noise average
\begin{equation}
	\frac{1}{\det \Omega} = 
	\langle \exp(-\xi^\dagger(\Omega - 1)\xi) \rangle_\xi.
	\label{eq:noise_avg}
\end{equation}
The reweighting factor is evaluated by using a Gaussian noise vector $|\xi\rangle$ as
\begin{align}
	|v_1\rangle &= f^{\rm poor}_{1/4}(\mathcal{D}(m_q)) f^{\rm poor}_{1/4}(\mathcal{D}(1)) f^{\rm poor}_{-1/4}(\mathcal{D}(1)) |\xi\rangle,  \nonumber \\
	|v_2\rangle &= f^{\rm poor}_{-1/2}\mathcal{D}(m_q) |v_1\rangle,  \nonumber \\
	\frac{1}{w} &= \exp[-(\omega - \xi_2)], \quad \omega \equiv \langle v_1 | v_2 \rangle, \quad \xi_2 \equiv \langle \xi | \xi \rangle.
\end{align}

\subsection{Results}
\begin{figure}[!htbp]
	\centering
	\includegraphics[scale=0.5]{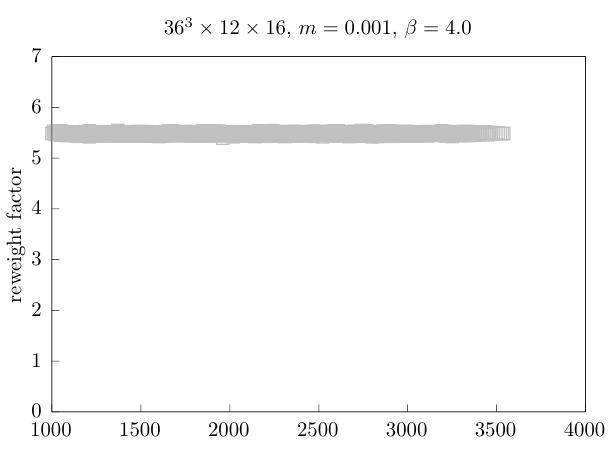}~
	\includegraphics[scale=0.5]{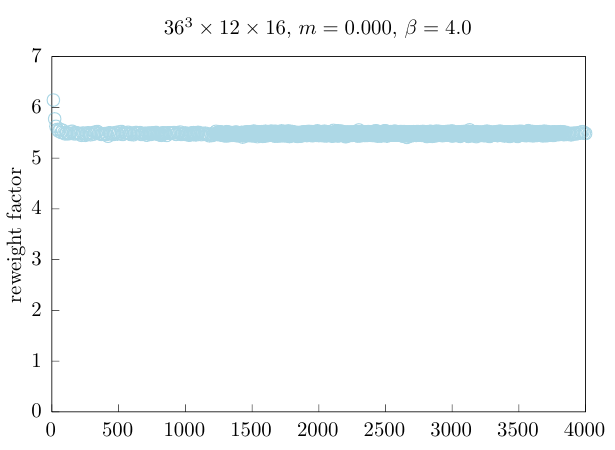}
	\includegraphics[scale=0.5]{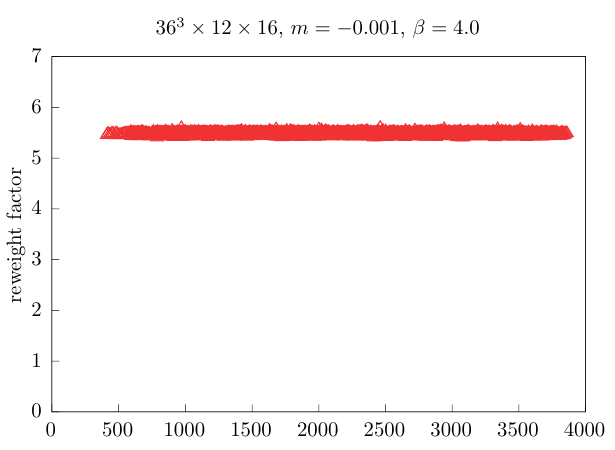}~
	\includegraphics[scale=0.5]{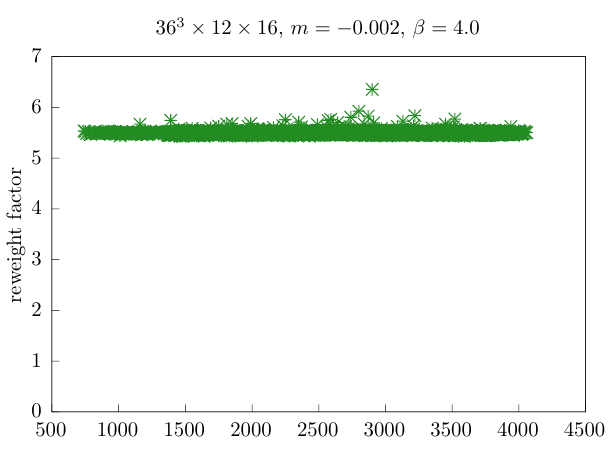}
		\includegraphics[scale=0.5]{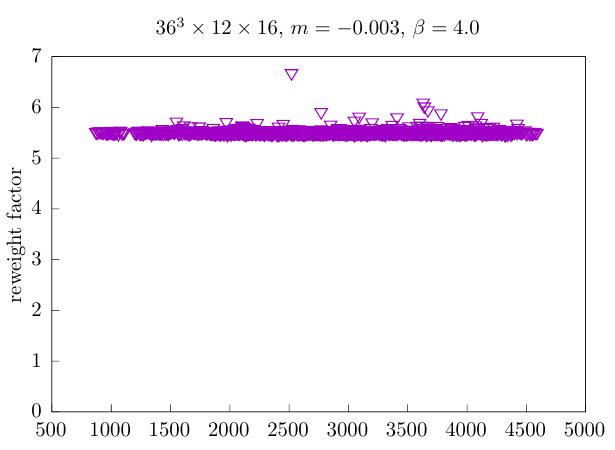}~
			\includegraphics[scale=0.5]{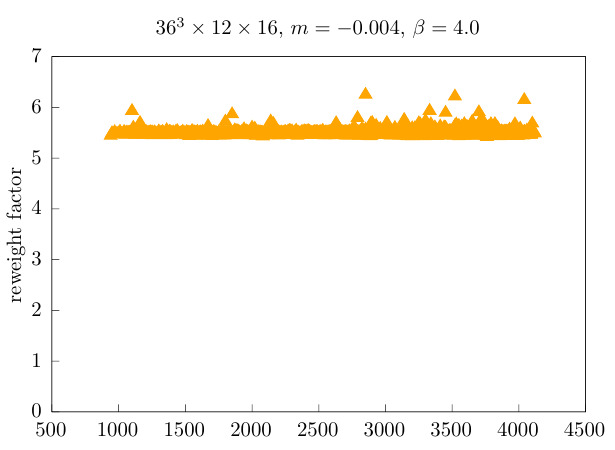}
\caption{The reweighting factors for the rational function 
	reweighting on the $36^3\times 12\times 16$ ensembles at 
	$\beta=4.0$, for quark masses $m_q = 0.001, 0.000, -0.001, 
	-0.002, -0.003, -0.004$. The reweighting factors are stable 
	throughout the Monte Carlo history with no large fluctuations, 
	indicating that the reweighting is well under control and can 
	be safely applied to correct for the poor rational function 
	approximation used in the original runs.}
	\label{fig:reweighting_factor}
\end{figure}

Fig.~\ref{fig:reweighting_factor} shows the reweighting factors for configurations generated with the earlier version of Grid. Their stability throughout the Monte Carlo history indicates that the reweighting can be safely applied.
We can also estimate the magnitude of the reweighting factors. By neglecting the difference between \(D_{\mathrm{MDWF}}(1)\) and \(D_{\mathrm{MDWF}}(m_q)\), the correction to the fermion action is
\begin{equation}
\int dx\,\rho(x)\left[ f_{1/4}^{\text{poor}}(x) f_{1/4}^{\text{poor}}(x) f_{1/4}^{\text{poor}}(x) f_{1/4}^{\text{poor}}(x) f_{1/2}^{\text{poor}}(x) f_{1/4}^{\text{poor}}(x) f_{1/4}^{\text{poor}}(x) - 1 \right]
\label{eq:poor-rational-integral} \end{equation}
where $\rho(x)$ is the spectral density of the matrix. Assuming that $\rho(x)$ is constant, integration over the range $10^{-4}\leq x\leq100$ gives $1.4\times10^{-8}$. Since the magnitude of the pseudofermion action is $5.37\times10^{7}$, the resulting correction to the action is estimated to be $1.4\times10^{-8}\times5.37\times10^{7}=0.75$, which provides an estimate of the logarithm of the reweighting factor. The actual value \(\sim\log 6 = 1.8\) is larger but of the same order of magnitude.

\begin{figure}[!htbp]
	\centering
\includegraphics[scale=0.65]{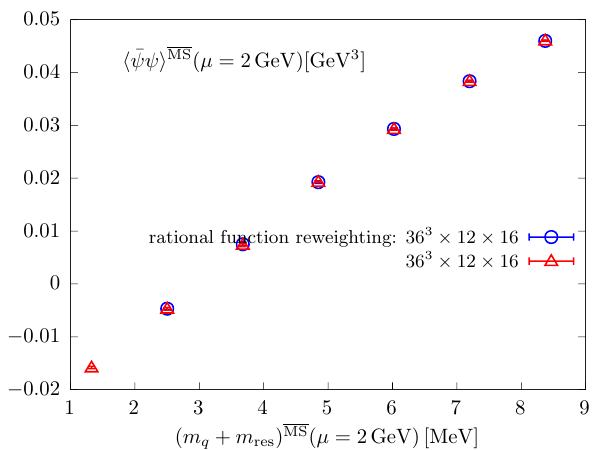}~
\includegraphics[scale=0.65]{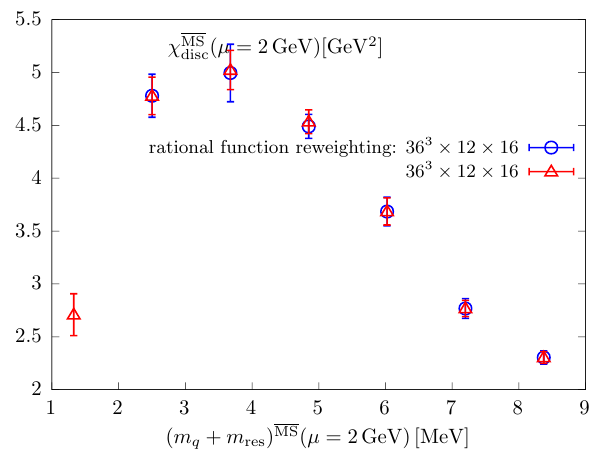}
\caption{Comparison of the chiral condensate 
			$\langle\bar{\psi}\psi\rangle^{\overline{\mathrm{MS}}}(\mu = 2\,\mathrm{GeV})$ 
			(left) and the disconnected chiral susceptibility 
			$\chi_\mathrm{disc}^{\overline{\mathrm{MS}}}(\mu = 2\,\mathrm{GeV})$ 
			(right) on the $36^3\times 12\times 16$, $\beta=4.0$ ensembles, 
			before (red triangles) and after (blue circles) applying the 
			rational function reweighting, as a function of the renormalized 
			quark mass $(m_q + m_\mathrm{res})^{\overline{\mathrm{MS}}}
			(\mu = 2\,\mathrm{GeV})$.} 
	\label{fig:compar_reweight_obs}
\end{figure}

Fig.~\ref{fig:compar_reweight_obs} shows the comparison of the chiral condensate
$\langle\bar{\psi}\psi\rangle^{\overline{\mathrm{MS}}}(\mu = 2\,\mathrm{GeV})$
and the disconnected chiral susceptibility
$\chi_\mathrm{disc}^{\overline{\mathrm{MS}}}(\mu = 2\,\mathrm{GeV})$ before and
after rational function reweighting on the $36^3\times 12\times 16$, $\beta=4.0$
ensembles. The two results are in good agreement within statistical errors, as
expected from Fig.~\ref{fig:reweighting_factor}, where the reweighting factors are
seen to remain stable along the Monte Carlo history. We have found no case in
which the reweighting changes a result beyond its statistical uncertainty; it is
nevertheless included throughout for consistency.
\clearpage
\section{Residual and Pion Screening Masses}
\label{app:additional mass}
\begin{figure}[!htbp]
	\centering
	\includegraphics[scale=0.65]{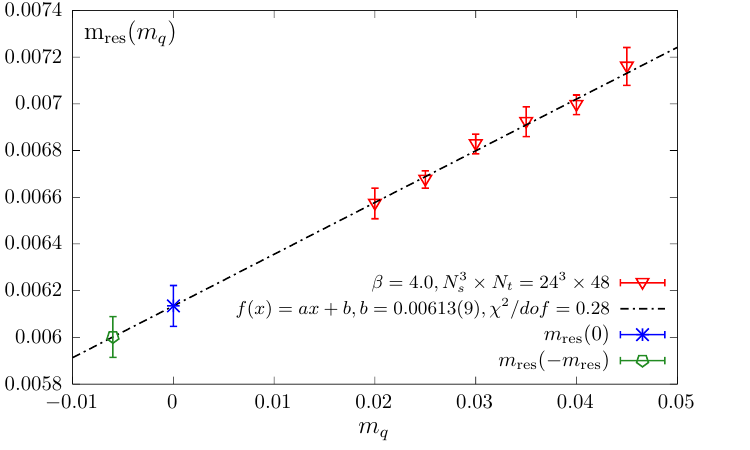}~	
	\includegraphics[scale=0.65]{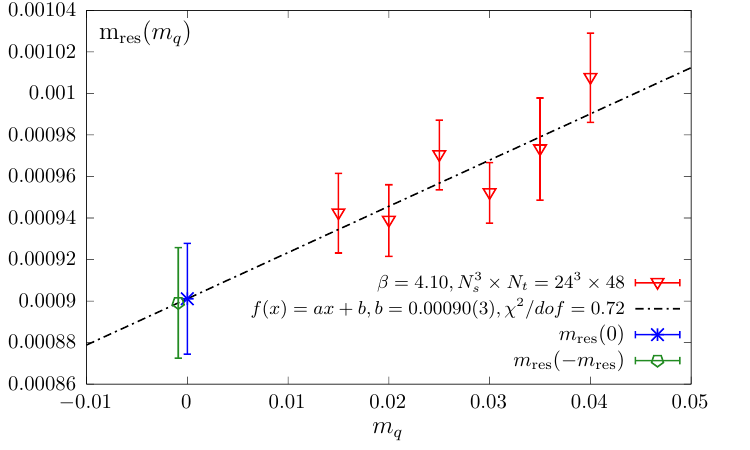}
\includegraphics[scale=0.65]{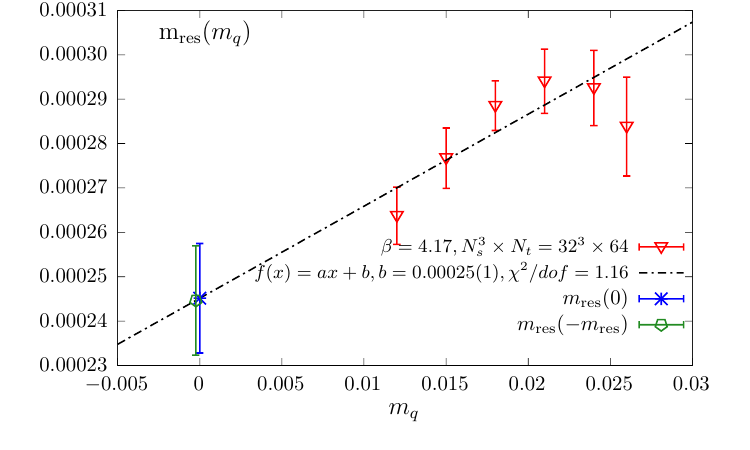}
	\caption{Residual mass $\mres$ as a function of the input quark mass, determined from the ratio of matrix elements of $J_{5q}$ and $J_5$. Results are shown for $\beta=4.0$ (left), $\beta = 4.10$ (right), and $\beta=4.17$ (bottom). Dash-dotted lines indicate linear fits. The residual masses extrapolated to bare input quark mass 0 and $-\mres$ are shown. The two results are consistent within errors for all three $\beta$ values. }
		\label{fig:sup_mres}
\end{figure}

The residual-mass extrapolations to the bare input quark masses $m_q=0$ and $m_q=-m_{\rm res}$ for three values of $\beta$ are shown in Fig.~\ref{fig:sup_mres}.

\begin{figure}[!htbp]
	\centering	\includegraphics[scale=0.5]{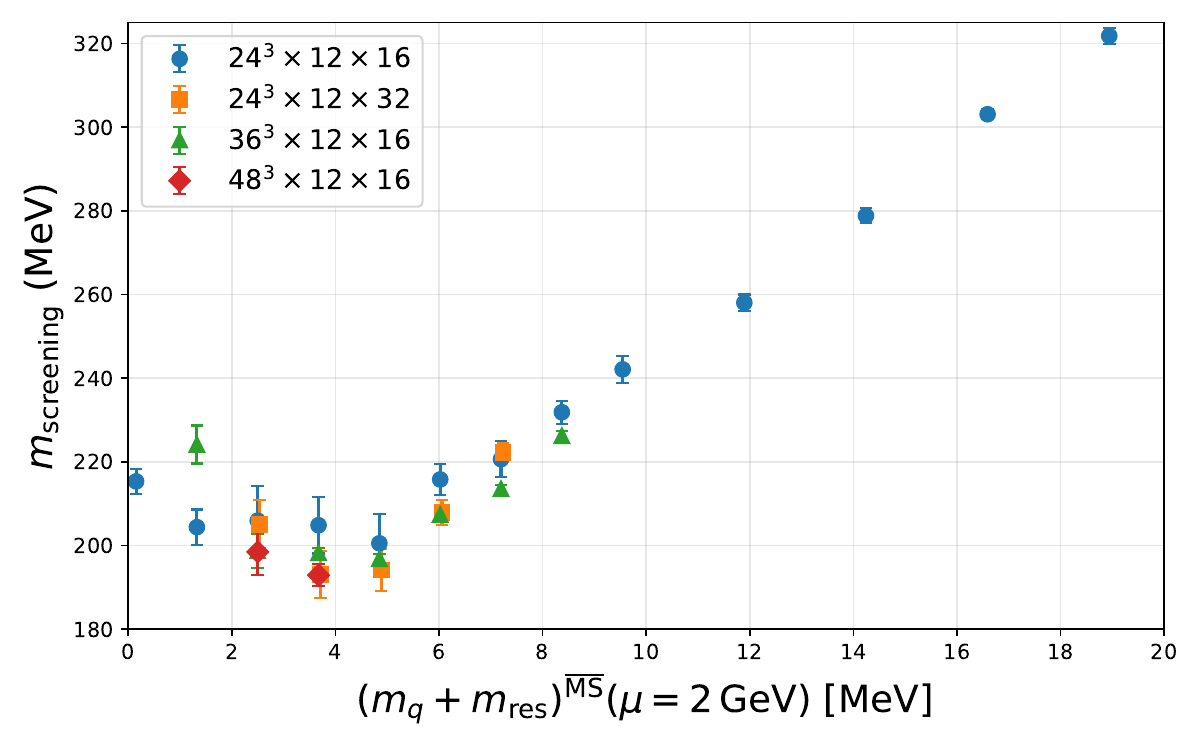}	
		\caption{The pion screening mass as a function of the renormalized quark mass for all $N_t=12$ lattices.}
	\label{fig:screening_mass}
\end{figure}

The pion screening masses used to fix the exponential scale in the
finite-volume fit are summarized in Fig.~\ref{fig:screening_mass}.

\clearpage
\bibliographystyle{apsrev4-2}
\bibliography{ref}
\end{document}